\documentclass[prb,showpacs,twocolumn,superscriptaddress,aps,a4paper]{revtex4}
\usepackage{dcolumn}
\usepackage{amsmath}
\usepackage{graphicx}
\usepackage{latexsym}
\usepackage{amsfonts}
\usepackage{amssymb}
\DeclareGraphicsExtensions{.pdf,.gif,.jpg}

\newcommand{\be}{\begin{equation}}
\newcommand{\ee}{\end{equation}}
\newcommand{\beq}{\begin{eqnarray}}
\newcommand{\eeq}{\end{eqnarray}}

\begin{document}



\title{Equilibrium and time-dependent Josephson current in one-dimensional
superconducting junctions}





\author{Enrico Perfetto}
\affiliation{Consorzio Nazionale Interuniversitario per le Scienze
Fisiche della Materia, Unit\'a Tor Vergata, Via della Ricerca
Scientifica 1, 00133 Rome, Italy}

\author{Gianluca Stefanucci}
\affiliation{Dipartimento di Fisica, Universit\'a di Roma Tor
Vergata, Via della Ricerca Scientifica 1, I-00133 Rome, Italy}
\affiliation{European Theoretical Spectroscopy Facility (ETSF)}

\author{Michele Cini}
\affiliation{Consorzio Nazionale Interuniversitario per le Scienze
Fisiche della Materia, Unit\'a Tor Vergata, Via della Ricerca
Scientifica 1, 00133 Rome, Italy}
\affiliation{Dipartimento di
Fisica, Universit\'a di Roma Tor Vergata, Via della Ricerca
Scientifica 1, I-00133 Rome, Italy}




\begin{abstract}
We investigate the transport properties of a one-dimensional
superconductor-normal metal-superconductor (S-N-S) system
described within the tight-binding approximation. We compute the
equilibrium dc Josephson current and the time-dependent
oscillating current generated after the switch-on of a constant
bias. In the first case an exact embedding procedure to calculate
the Nambu-Gorkov Keldysh Green's function is employed and used to
derive the continuum and bound states contributions to the dc
current. A general formalism to obtain the Andreev bound states
(ABS) of a normal chain connected to superconducting leads is also
presented. We identify a regime in which all Josephson current is
carried by the ABS and obtain an analytic formula for the
current-phase relation in the limit of long chains. In the latter
case the condition for perfect Andreev reflections is expressed in
terms of the microscopic parameters of the model, showing a
limitation of the so called wide-band-limit (WBL) approximation.
When a finite bias is applied to the S-N-S junction we compute the
exact time-evolution of the system by solving numerically the
time-dependent Bogoliubov-deGennes equations. We provide a
microscopic description of the electron dynamics not only inside
the normal region but also in the superconductors, thus gaining
more information with respect to WBL-based approaches. Our scheme
allows us to study the ac regime as well as the transient dynamics
whose characteristic time-scale is dictated by the velocity of
multiple Andreev reflections.
\end{abstract}

\maketitle


\section{Introduction}

In the last few years nanoscopic Josephson junctions have been
widely studied both theoretically and experimentally as possible
candidates to provide an alternative technology to silicon-based
electronics\cite{review,martins,likharev,sset1,sset2,amplifier}.
Special attention has been paid to the analysis of superconducting
atomic-size quantum point contacts (SQPC)\cite{sqpc} like
single-level quantum dots and nanowires. Among the most striking
features experimentally observed we mention the subgap structure
in the current-voltage characteristics driven by multiple Andreev
reflections\cite{muller}, the single-electron tunneling through
discrete electronic states\cite{ralph}, and the nanoscopic dc
Josephson current\cite{koops}.

Within the so-called \textit{Hamiltonian approach}\cite{cuevas1}
it is possible to provide an accurate microscopic description of
these systems, where some relevant length scales (Fermi length,
size of the junction, etc.) are comparable. This approach relies
on tight-binding-like Hamiltonians and has the advantage to treat
the tunneling Hamiltonian describing the SQPC to all
orders\cite{rodero,cuevas1}. In SQPC the Andreev bound states
(ABS)\cite{andreev,beenakker} play an important role since they
can carry an important amount of dc Josephson
current\cite{kulik,beenakker,furusaki}. Such states origin from
multiple Andreev reflections occurring at the
superconductor-device contact and come in pairs, one above and one
below the Fermi level, carrying opposite supercurrents. In spite
of the large theoretical effort in studying the dc Josephson
regime in SQPC, a proper description of extended junctions is
still lacking since the electrodes degrees of freedom have been so
far absorbed in an approximate frequency-independent pairing and
on-site potentials at the boundaries of the central
region\cite{affleck2,feiguin,bergeret}.

The calculation of the ac Josephson current is more involved. At
present the ac regime has been studied using Floquet-based methods
combined with nonequilibrium Green's function
techniques\cite{cuevas1,cuevas2,guo}. This approach, however, is
limited to the dc bias case and other interesting time-dependent
driving fields, like ac bias or voltage pulses, cannot be
addressed. A possible alternative approach is the one based on the
real time-propagation but, so far, only normal metal-quantum
dot-superconductor junctions have been studied\cite{tdtheor1}.

In this paper we investigate the transport properties of a
one-dimensional (1D) superconductor-normal metal-superconductor
(S-N-S) system\cite{nota} composed by a normal tight-binding chain
embedded between two 1D superconductors described by the
Bogoliubov-deGennes Hamiltonian. We will study both the static dc
Josephson current $J$ and the time-dependent oscillating current
generated after the switch-on of a constant bias. In the dc case
we employ an exact embedding procedure and calculate the three
different contributions to $J$, carried by the ABS, the normal
bound states (with energy below the bottom of the band), and the
continuum states. We show that if the pairing potential is larger
than half the bandwidth of the normal region, all Josephson
current is carried by the ABS's. In this regime we are able to
extend the results by Affleck et al.\cite{affleck2} in the limit
of long normal region. The use of the exact embedding self-energy
allows us to relate the phenomenological paring potential of
Ref.\onlinecite{affleck2} with the microscopic parameters of the
model, thus obtaining a condition for perfect Andreev reflections
in term of the physical order parameter $\Delta$. In addition we
highlight a limitation of the commonly used wide-band-limit (WBL)
approximation.

When a finite bias is applied to the S-N-S junction, we compute
the exact time-evolution of the system by solving numerically the
time-dependent Bogoliubov-deGennes
equations\cite{andreev,degennes2,degennes3}. This is done within
the so-called partition-free approach, in which the S-N-S system
is assumed to be contacted and in equilibrium before the external
bias is switched on.\cite{cini,stefanucci}. Explicit calculations
are performed in the case of superconducting leads of finite
length. However, as already discussed in Ref.\onlinecite{tdspin},
the electrodes are long enough to reproduce the time evolution of
the infinite-leads system. The above approach gives us the
possibility to explore the transient dynamics and provides a
time-dependent picture of the Andreev reflections. In the
long-time limit we recover the expected oscillating current, whose
Fourier transform displays contributions from different harmonics
of the fundamental Josephson frequency. By extracting the dc
component of the oscillating current, we are also able to
reproduce the subgap structure in the current-voltage
characteristics.

The paper is organized as follows. In Section \ref{model} we
introduce the model Hamiltonian and briefly recall the Nambu and
Bogoliubov-deGennes formalisms. In Section \ref{jossect} the
equilibrium Josephson current is studied by means of an exact
embedding procedure.  Numerical results for short junctions are
reported in Section \ref{numerical} while the limit of long normal
regions is analytically carried out in Section \ref{seclong}. In
Section \ref{tdsec} we investigate the time-dependent regime. Two
Appendices corroborate the analytic derivations. Finally summary
and main conclusions are drawn in Section \ref{conclusions}.

\section{The model}
\label{model}

\begin{figure}[h]
\includegraphics[height=2.3cm ]{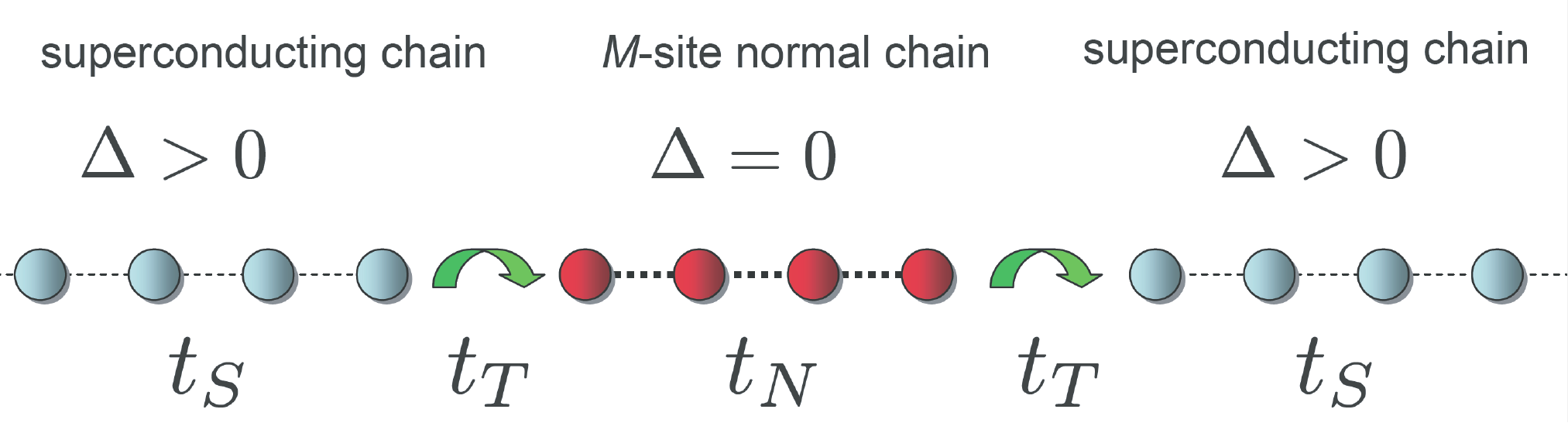}
\caption{Scheme of the S-N-S junction. For illustration the
superconducting leads are 1D chain with nearest-neighbor hopping
$t_{S}$ [i.e. $\varepsilon_{q}=2t_{S}\cos q$ in Eq.(\ref{hlead})]
and on-site pairing potentials $\Delta_{L}=\Delta_{R}=\Delta$ with
$\chi_{L}=\chi_{R}=0$. The hopping integral between the boundary
sites of the superconducting and normal regions is $t_{T}$ [i.e.
$V_{q}=t_{T}\sin q \sqrt{2/\Lambda}$ in Eq.(\ref{htunn}), where
$\Lambda$ is the number of sites in the leads].} \label{device}
\end{figure}

We consider a hybrid S-N-S system consisting of a normal region
contacted to two superconductors, as illustrated in
Fig.\ref{device}. In the Bogoliubov-deGennes formalism the
annihilation (creation) fermion operators
$c_{\uparrow}^{(\dagger)}$ annihilates (creates)
\textit{electrons} of spin up, while the annihilation (creation)
fermion operators $\tilde{c}^{(\dagger)}_{\downarrow}$ annihilates
(creates) \textit{holes} of spin down. In order to avoid confusion
we put a tilde on the hole-operators. The Hamiltonian of the
system is described by
\begin{equation}
\hat{H}=\hat{H}_{N}+\hat{H}_{L}+\hat{H}_{R}+\hat{H}_{T}-\mu
(\hat{N}_{\uparrow}-\hat{N}_{\downarrow}) \,.
\label{ham}
\end{equation}
In this work we consider normal regions consisting of a
tight-binding chain of length $M$ and nearest neighbor hopping
$t_{N}$ with Hamiltonian
\begin{equation}
\hat{H}_{N}=t_{N}\sum_{i=1}^{M-1}
\left[c^{\dagger}_{i\uparrow}c_{i+1\uparrow}
-\tilde{c}^{\dagger}_{i\downarrow}\tilde{c}_{i+1\downarrow}\right]
+ \mathrm{h.c.} \,.
\end{equation}
The Hamiltonians for the Left/Right (L/R) superconducting leads
has the general form
\begin{eqnarray}
\hat{H}_{\alpha}&=&\sum_{q} \left[ \varepsilon_{q}
c^{\dagger}_{q\alpha \uparrow}c_{q \alpha
\uparrow}-\varepsilon_{q} \tilde{c}^{\dagger}_{q \alpha
\downarrow}
\tilde{c}_{q \alpha \downarrow} \right. \nonumber \\
 &+& \left.  \Delta_{\alpha}e^{i\chi_{\alpha}} c^{\dagger}_{q \alpha
\uparrow}\tilde{c}_{q \alpha
\downarrow}+\Delta_{\alpha}e^{-i\chi_{\alpha}}
\tilde{c}^{\dagger}_{q \alpha \downarrow}c_{q \alpha \uparrow}
\right] \,,
\label{hlead}
\end{eqnarray}
where $\Delta_{\alpha}$ is the pairing potential in lead
$\alpha=L,R$ with corresponding phase $\chi_{\alpha}$. The
one-particle energies $\varepsilon_{q}$ span the range $(-W,W)$
where $2W$ is the lead bandwidth. We assume that the tunneling
between the superconductors and the normal region occurs only via
the boundary sites of the chain and model $H_{T}$ as
\begin{eqnarray}
\hat{H}_{T}&=&\sum_{q} V_{q} \left[ c^{\dagger}_{q L \uparrow}
c_{1 \uparrow}+
c^{\dagger}_{q R \uparrow} c_{M \uparrow} \right. \nonumber \\
&-& \left. \tilde{c}^{\dagger}_{q L \downarrow} \tilde{c}_{1
\downarrow}- \tilde{c}^{\dagger}_{q R \downarrow} \tilde{c}_{M
\downarrow} \right] + \mathrm{h.c.} \,.
\label{htunn}
\end{eqnarray}
In the last term of Eq.(\ref{ham}) $\mu$ is the chemical potential
and $\hat{N}_{\uparrow / \downarrow}$ is the number of
electron/holes with spin $\uparrow /\downarrow$.

The time-dependent current\cite{nota2} at the $\alpha = L,R $
interface is
\begin{equation}
I_{\alpha}(t)=   2 \sum_{q}V_{q} \mathrm{Re}\, \mathrm{Tr}
\,\left[ \underline{G}^{<}_{1,q \alpha}(t,t) \, \right] \,,
\label{current}
\end{equation}
where the Nambu lesser Green's function is defined as
\begin{eqnarray}
&&[\underline{\mathbf{G}}^{<}(t_{1},t_{2})]_{m,n}\equiv
\underline{G}^{<}_{m,n}(t_{1},t_{2}) \nonumber \\
&& =i
 \left(\begin{array}{ll}
 \langle
c^{\dagger}_{m\uparrow}(t_{1})c_{n\uparrow}(t_{2})\rangle& \langle
\tilde{c}^{\dagger}_{m\downarrow}(t_{1})c_{n\uparrow}(t_{2})\rangle  \\
 \langle
c^{\dagger}_{m\uparrow}(t_{1})\tilde{c}_{n\downarrow}(t_{2})\rangle
& \langle
\tilde{c}^{\dagger}_{m\downarrow}(t_{1})\tilde{c}_{n\downarrow}(t_{2})\rangle
\end{array}\right) \, .
\label{gless1}
\end{eqnarray}
In the above definition the indices $m,n$ denote either a site in
the normal chain or a $q$-state in the $\alpha=L,R$ lead. We
observe that the off-diagonal components of the Green's function
can be interpreted as \textit{spin-flip} propagators in the
effective Bogoliubov-deGennes space. The retarded, advanced and
greater Green's functions are defined in a similar way as in
Eq.(\ref{gless1}).

The rest of the paper is devoted to the calculation of
$I_{\alpha}(t)$. First we will focus on the equilibrium problem
and calculate the \textit{dc Josephson current}
$J=I_{L}(0)=I_{R}(0)$. Then we apply a finite bias voltage across
the junction and compute numerically the time-dependent current
$I_{L}(t)$ at the left interface.

\section{dc Josephson current}
\label{jossect}

The dc Josephson current $J=I_{L}(0)$ is obtained from
Eq.(\ref{current}) with an equilibrium lesser Green's function and
reads
\begin{equation}
J=  2 \mathrm{Re}\, \int \frac{d\omega}{2\pi}
 \mathrm{Tr} \,\left[ \sum_{q} V_{q}
\underline{G}^{<}_{1,Lq}(\omega) \right] \,. \label{currentdc}
\end{equation}
In equilibrium the lesser Green's function is related to the
retarded/advanced Green's function via the fluctuation-dissipation
theorem
\begin{equation}
\underline{\mathbf{G}}^{<}(\omega) = -f(\omega)
\left[\underline{\mathbf{G}}^{r}(\omega)-
\underline{\mathbf{G}}^{a}(\omega) \right] \, ,
\end{equation}
where $f$ is the Fermi distribution function. In the following we
work at zero temperature. This means that the effective pairing
potential in Eq.(\ref{hlead}) corresponds to the BCS gap at $T=0$.
The entire formalism remains valid at finite temperature $T$,
provided that the order parameter $\Delta$ corresponds to the BCS
gap at $T$. The dependence on temperature of the current $J$ is
mainly due to the change of $\Delta(T)$, since the Fermi function
$f$ remains close to a theta function for $T \lesssim \Delta$.
This is supported by the results shown in Fig.\ref{cross} which
agree well with previous studies on temperature dependence of $J$
\cite{kummel1}.

By exploiting the Dyson equation for the retarded/advanced Green's
function the Josephson current $J$ can be expressed in terms of
the embedding self-energy
\begin{equation}
\underline{\Sigma}^{r/a}_{\alpha}(\omega)=\sum_{q} V_{q}^{2}  \,
\sigma_{z} \, \underline{g}^{r/a}_{\alpha q} (\omega) \,
\sigma_{z}
\label{selfen}
\end{equation}
as
\begin{equation}
J=  2  \mathrm{Re}   \int \frac{d\omega}{2\pi}  \mathrm{Tr} \,
\left\{ \left[ \underline{G}^{r}_{1,1} (\omega) \,
\underline{\Sigma}^{r}_{L}(\omega) - \underline{G}^{a}_{1,1}
(\omega) \, \underline{\Sigma}^{a}_{L}(\omega)  \right] \sigma_{z}
  \right\}  ,
\label{joscurr}
\end{equation}
where $\sigma_{z}$ is the third Pauli matrix and
$\underline{g}^{r/a}_{\alpha q}$ is the Green's function of the
isolated $\alpha$ lead. We observe that Eq.(\ref{joscurr}) is
valid for any S-N-S system provided that the S-N hopping occurs
only at the two boundary sites of the normal region. The general
expression for the embedding self energy is
\begin{equation}
\underline{\Sigma}^{r/a}_{\alpha}(\omega)=\left(
\begin{array}{cc}m_{\alpha}(\omega \pm i\eta)&
\tilde{\Delta}_{\alpha}(\omega \pm i\eta) e^{i\chi_{\alpha}}
\\\tilde{\Delta}_{\alpha}(\omega \pm i\eta)
e^{-i\chi_{\alpha}}&m_{\alpha}(\omega \pm i\eta)
\end{array}\right) \, ,
\label{selfenergymat}
\end{equation}
where $m$ and $\tilde{\Delta}$ are the effective on-site and
pairing potentials. In the case of 1D superconducting leads with
$\Lambda$ sites (see Fig.\ref{device})
\begin{equation}
\varepsilon_{q} = 2t_{S} \cos q \quad , \quad V_{q}=t_{T}
\sqrt{\frac{2}{\Lambda}} \sin q \label{1dcoupling}
\end{equation}
and the self-energy at $\mu = 0$ is (see Appendix \ref{self})
\begin{eqnarray}
m_{\alpha}^{\mathrm{1D}}(z) = z \frac{t_{T}^{2}}{2t_{S}^{2}}
\frac{\sqrt{\Delta_{\alpha}^{2}-z^{2}}-\sqrt{\Delta_{\alpha}^{2}-
z^{2}+4t_{S}^{2}}}{\sqrt{\Delta_{\alpha}^{2}-z^{2}}} ,
\label{embedding1}
\\
\tilde{\Delta}_{\alpha}^{\mathrm{1D}}(z)=\Delta_{\alpha}
\frac{t_{T}^{2}}{2t_{S}^{2}} \frac{\sqrt{
z^{2}-\Delta_{\alpha}^{2}-4t_{S}^{2}}-\sqrt{z^{2}-
\Delta_{\alpha}^{2}}}{\sqrt{z^{2}-\Delta_{\alpha}^{2}}},
\label{embedding2}
\end{eqnarray}
where $z$ is a complex frequency and the infinite $\Lambda$ limit
has been taken. The WBL result is easily recovered by defining the
tunneling rate $\Gamma=2t_{T}^{2}/t_{S}$, expanding
Eqs.(\ref{embedding1},\ref{embedding2}) in powers of $z/t_{S}$ and
$\Delta/t_{S}$ and retaining only the zero-th order term. In this
way one gets
\begin{eqnarray}
m^{\rm{WBL}}_{\alpha}(z) &=& -\frac{\Gamma}{2}
\frac{z}{\sqrt{\Delta_{\alpha}^{2}-z^{2}}} \, ,
\nonumber \\
\tilde{\Delta}^{\rm{WBL}}_{\alpha}(z)&=&\frac{\Gamma}{2}
\frac{\Delta_{\alpha}}{\sqrt{\Delta_{\alpha}^{2}-z^{2}}} \,,
\label{embeddingwbl} \nonumber
\end{eqnarray}
which for $z = \omega + i\eta$ yields the commonly used WBL
self-energy $\Sigma^{\rm{WBL}}$. We would like to observe that
evaluating $\Sigma^{\rm{WBL}}$ at the Fermi energy $\omega=\mu=0$
one finds that $m^{\rm{WBL}}=0$ and that the pairing potential
$\tilde{\Delta}^{\rm{WBL}} \propto \Gamma$ is independent of the
order parameter $\Delta$. In Section \ref{seclong} we will discuss
the implications of this feature for long normal chains.

\begin{figure}[h]
\includegraphics[height=4.cm ]{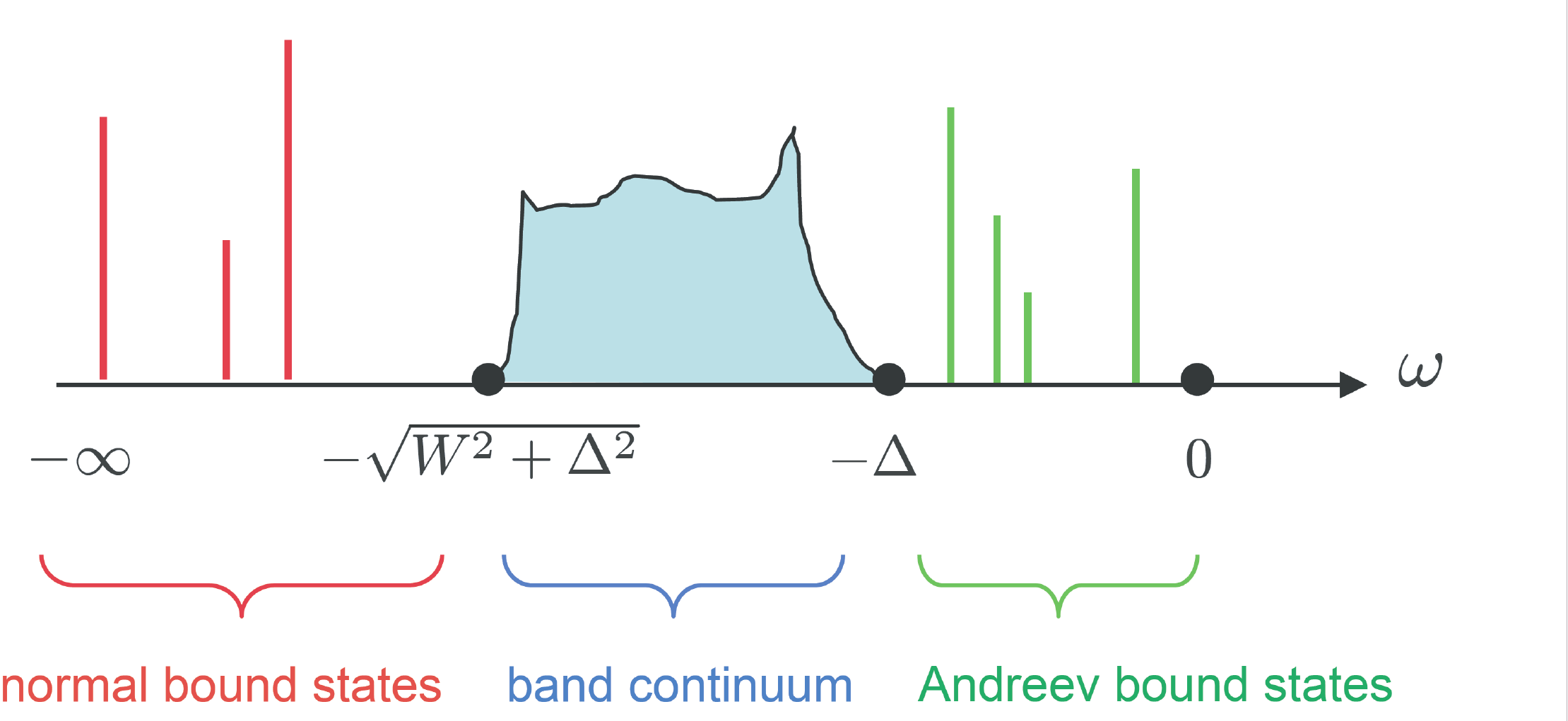}
\caption{Schematic representation of the density of states of the
S-N-S system. The three spectral regions corresponding to normal
bound states, Andreev bound states and continuum states are
displayed assuming $\Delta_{L}=\Delta_{R}=\Delta$.}
\label{spectral}
\end{figure}

\begin{figure}[h]
\includegraphics[height=4.5cm ]{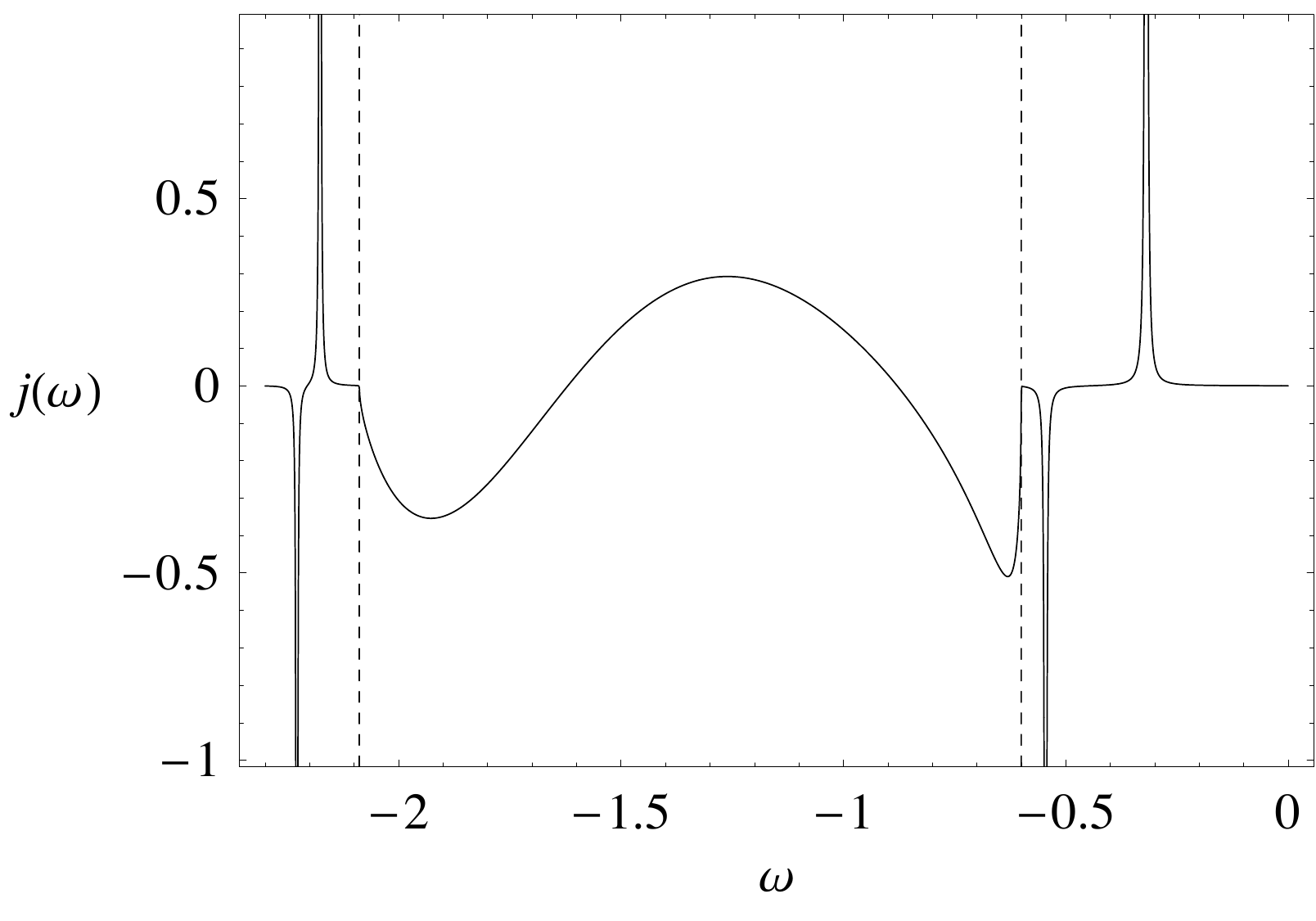}
\caption{Integrand function $j(\omega)$ in Eq.(\ref{joscurr}) for
a 1D S-N-S junction with $M=4$, $t_{S}=t_{T}=1$, $t_{N}=1.2$,
$\Delta_{L}=\Delta_{R}=0.6$, $\chi=\chi_{L}-\chi_{R}=\pi/3$. The
two dashed vertical lines correspond to
$\omega=-\sqrt{W^{2}+\Delta_{\rm{max}}^{2}}$ and
$\omega=-\Delta_{\rm{min}}$ and mark the boundaries of the three
integration regions. A broadening $\eta=10^{-4}$ has been used to
give to $j$ a finite width around the bound states. Energies are
in units of $|t_{S}|$.} \label{spettro}
\end{figure}

In the rest of the Section we do not assume any specific form of
the embedding self-energy and present a general procedure to
calculate the dc Josephson current of Eq.(\ref{joscurr}). For
practical purposes we split the integral in Eq.(\ref{joscurr}) in
three different energy regions and identify the contributions of
the \textit{normal bound states}, \textit{Andreev bound states}
and \textit{continuum states} (see Fig.\ref{spectral}).
The energy range is
$(-\sqrt{W^{2}+\Delta_{\rm{max}}^{2}},-\Delta_{\rm{min}})$ for the
filled continuum states, $(-\Delta_{\rm{min}},0)$ for the filled
ABS's and $(-\infty, -\sqrt{W^{2}+\Delta_{\rm{max}}^{2}})$ for the
filled normal bound states, where
$\Delta_{\rm{max}}=\max\{\Delta_{L},\Delta_{R}\}$ and
$\Delta_{\rm{min}}=\min\{\Delta_{L},\Delta_{R}\}$. Thus letting
$j(\omega)$ be the integrand function in Eq.(\ref{joscurr}) the
total Josephson current reads
\begin{eqnarray}
J&=&J_{\rm{cont}}+J_{\rm{abs}}+J_{\rm{nbs}} \, ,  \\
J_{\rm{cont}} &=&
\int_{-\sqrt{W^{2}+\Delta_{\rm{max}}^{2}}}^{-\Delta_{\rm{min}}}
\frac{d\omega}{2\pi} \, j(\omega) \, ,
 \\
J_{\rm{abs}} &=& \int^{0}_{-\Delta_{\rm{min}}}
\frac{d\omega}{2\pi} \,j(\omega) \, ,
\label{jabs} \\
J_{\rm{nbs}} &=&
\int^{-\sqrt{W^{2}+\Delta_{\rm{max}}^{2}}}_{-\infty}
\frac{d\omega}{2\pi} \, j(\omega) \, . \label{jbbs}
\end{eqnarray}
The nature of the above decomposition  is illustrated in
Fig.\ref{spettro}, where the the integrand function $j(\omega)$ is
displayed for a 1D S-N-S junction at a fixed value of
$\chi=\chi_{L}-\chi_{R}=\pi/3$. In Fig.\ref{spettro} we have
chosen the superconducting gap $\Delta$ about one order of
magnitude smaller than the leads bandwidth $W=4t_{S}$. This is
done in order to highlight the contribution coming from the normal
bound states, although we expect that it becomes less and less
important as $\Delta/W \rightarrow 0$. We would like to emphasize,
however, that the normal bound states play a crucial role in a
self-consistent calculation of the total current, since the
effective mean-field potentials depend on the density and in the
central region the contribution of the normal bound states is
certainly not negligible. Indeed the total number of particles
$N_{N}$ in the central normal region is
\begin{equation}
N_{N}=  -i\int \frac{d\omega}{2\pi} \, \mathrm{Tr}  \,
\sum_{m=1}^{M}  [\underline{\mathbf{G}}^{<}(\omega)]_{m,m}  \, ,
\end{equation}
where the integrand function in the above equation has a similar
structure as in Fig.\ref{spettro}.

It is worth noticing that the function $j(\omega)$ is proportional
to a Dirac delta around the bound states. Therefore the numerical
integrals in Eqs.(\ref{jabs}) and (\ref{jbbs})  must be computed
with care. An efficient alternative way to calculate with high
accuracy $J_{\rm{abs}}$ and $J_{\rm{nbs}}$ consists in realizing
that
\begin{equation}
J_{\rm{abs}}=2\sum_{n}\frac{dE^{(n)}_{\rm{abs}}}{d\chi} \quad ,
\quad  J_{\rm{nbs}} = 2\sum_{m}\frac{dE^{(m)}_{\rm{nbs}}}{d\chi}
\, , \label{hell}
\end{equation}
where $E^{(n)}_{\rm{abs}}$ and $E^{(m)}_{\rm{nbs}}$ are the
energies of the filled Andreev and normal bound states
respectively. Eq.(\ref{hell}) follows directly from the
Hellmann-Feynmann theorem, which in this case can be exploited
since $E^{(n)}_{\rm{abs}}$ and $E^{(n)}_{\rm{nbs}}$ are the
eigenenergies of the Hamiltonian in Eq.(\ref{ham}). By a simple
gauge transformation the phase $\chi$ can be transferred to the
hopping integrals $V_{q}$ in Eq.(\ref{htunn}), and hence the
derivative of $E^{(n)}_{\rm{abs}}$ and $E^{(n)}_{\rm{nbs}}$ with
respect to $\chi$ yields the average of the current operator over
the Andreev and bound eigenstates. In the following we derive and
elegant formula to calculate $E^{(n)}_{\rm{abs}}$ and
$E^{(m)}_{\rm{nbs}}$.

The bound state energies can be obtained by solving the
self-consistent $2M \times 2M$ secular problem
\begin{equation}
\hat{H}_{N}^{\mathrm{eff}}(E)|\psi _{E}\rangle= E |\psi_{E}\rangle
\label{secular}
\end{equation}
with $|E|<\Delta_{\rm{min}}$ (Andreev) and $
|E|>\sqrt{W^{2}+\Delta_{\rm{max}}^{2}}$ (normal), and
\begin{eqnarray}
\hat{H}_{N}^{\mathrm{eff}}(E)&=&\hat{H}_{N} \nonumber \\
&+& m_{R}(E)\,[c^{\dagger}_{M\uparrow}c_{M\uparrow}+
\tilde{c}^{\dagger}_{M\downarrow}\tilde{c}_{M\downarrow}]
\nonumber \\
&+& \tilde{\Delta}_{R}(E)\,[e^{i\chi_{R}}c^{\dagger}_{M
\uparrow}\tilde{c}_{M\downarrow}+
e^{-i\chi_{R}}\tilde{c}^{\dagger}_{M
\downarrow}c_{M\uparrow}]\nonumber \\
 &+&m_{L}(E)\,
[c^{\dagger}_{1\uparrow}c_{1\uparrow}+
\tilde{c}^{\dagger}_{1\downarrow}\tilde{c}_{1\downarrow}]
\nonumber \\
&+&\tilde{\Delta}_{L}(E)\,[e^{i\chi_{L}}c^{\dagger}_{1
\uparrow}\tilde{c}_{1\downarrow}+
e^{-i\chi_{L}}\tilde{c}^{\dagger}_{1 \downarrow}c_{1\uparrow}],
\label{heff}
\end{eqnarray}
with $m$ and $\tilde{\Delta}$ as in Eq.(\ref{selfenergymat}). In
the effective Hamiltonian the on-site and pairing potentials at
the boundary sites 1 and $M$ are renormalized by the embedding
procedure. In order to simplify the algebra we define the momenta
$k$ such that $E=2t_{N}\cos(k)$ and assume
$\Delta_{L}=\Delta_{R}=\Delta_{\rm{max}}=\Delta_{\rm{min}}\equiv
\Delta$, which also implies
$\tilde{\Delta}_{L}(E)=\tilde{\Delta}_{R}(E)\equiv\tilde{\Delta}_{k}$
and $m_{L}(E)=m_{R}(E)\equiv m_{k}$. In Appendix \ref{boundstates}
we describe in detail how the eigenvalue problem in
Eq.(\ref{secular}) is analytically solved to yield the following
equation for the momenta $k$
\begin{eqnarray}
0&=& t_{N}^{4} \sin ^{2}k(M+1)
+(-1)^{M}2t_{N}^{2}\tilde{\Delta}_{k}^{2} \cos \chi \sin ^{2}k
\nonumber \\
&-&2t_{N}\tilde{\Delta}_{k}^{2} \sin^{2}(kM) +
\tilde{\Delta}_{k}^{4} \sin^{2}k(M-1) \nonumber
\\
&-&m_{k}^{2}[2t_{N}\sin (kM)-m_{k}\sin k (M-1)]^{2} \,,
\label{andreevt}
\end{eqnarray}
where $\chi=\chi_{L}-\chi_{R}$. The bound state energies are found
by solving Eq.(\ref{andreevt}) and retaining only the values of
$k$ for which $|2t_{N}\cos k | < \Delta$  and $|2t_{N}\cos k
|>\sqrt{W^{2}+\Delta^{2}}$. We observe that in general the
variable $k$ is \textit{complex}, see Appendix \ref{boundstates}.

We would like to end this Section by commenting two limiting
cases. For an isolated normal region ($m_{k}=\tilde
{\Delta}_{k}=0$), the allowed momenta are simply $k=\pi j/(M+1)$,
$j=1,...M$, as expected. We also observe that if we set $m_{k}=0$
and assume a constant pairing potential $\tilde
{\Delta}_{k}=\tilde {\Delta}$, the above equation reduces to
Eq.(3.3) of Ref.\onlinecite{affleck2}.

\section{Numerical results for $J(\chi)$}
\label{numerical}

\begin{figure}[h]
\includegraphics[height=2.9cm ]{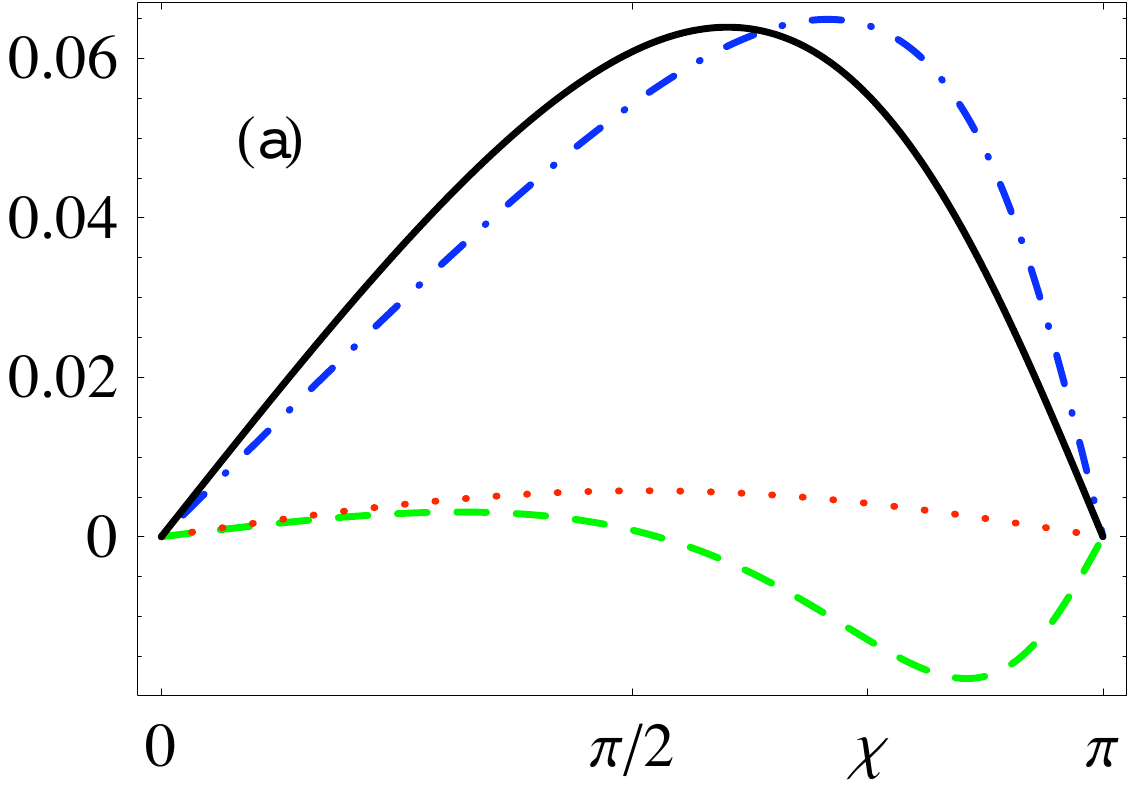}
\includegraphics[height=2.9cm ]{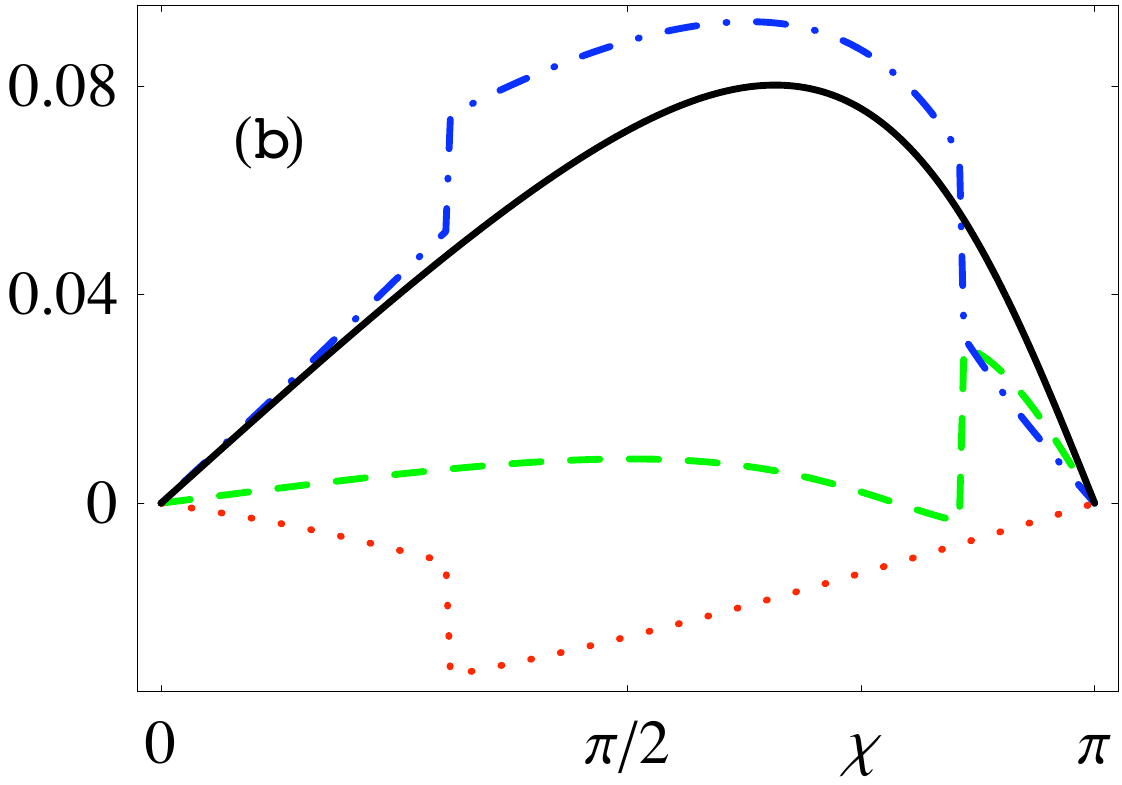}\\
\includegraphics[height=2.9cm ]{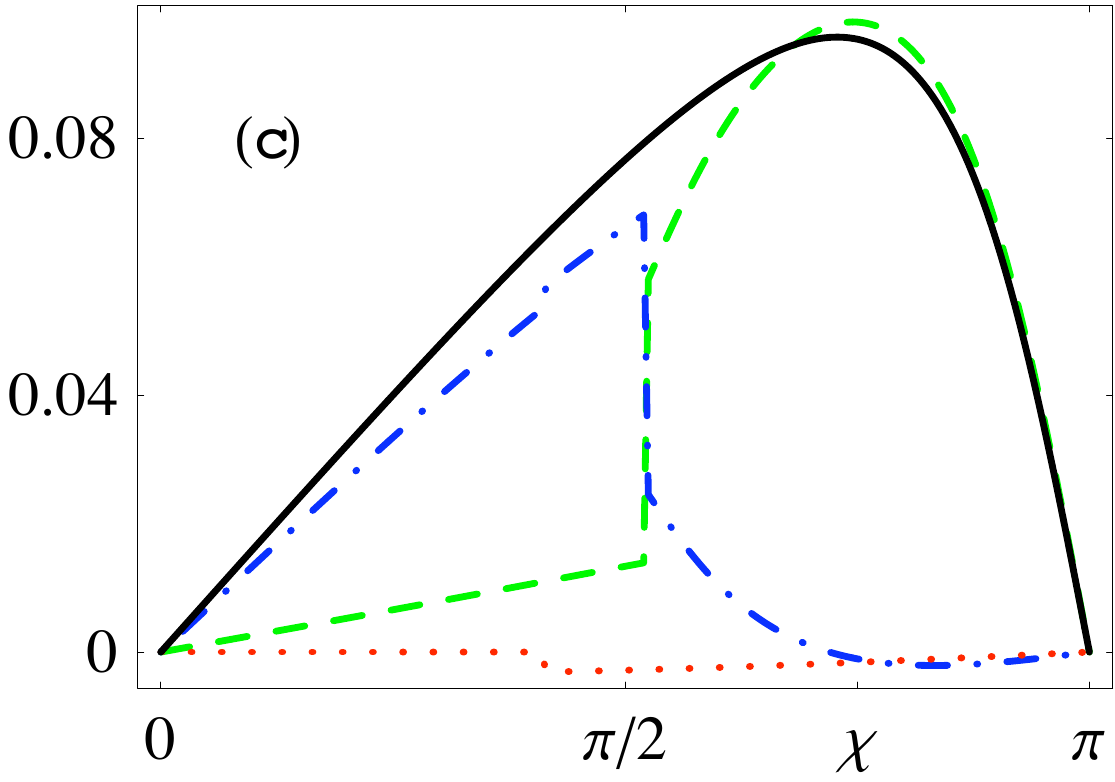}
\includegraphics[height=2.9cm ]{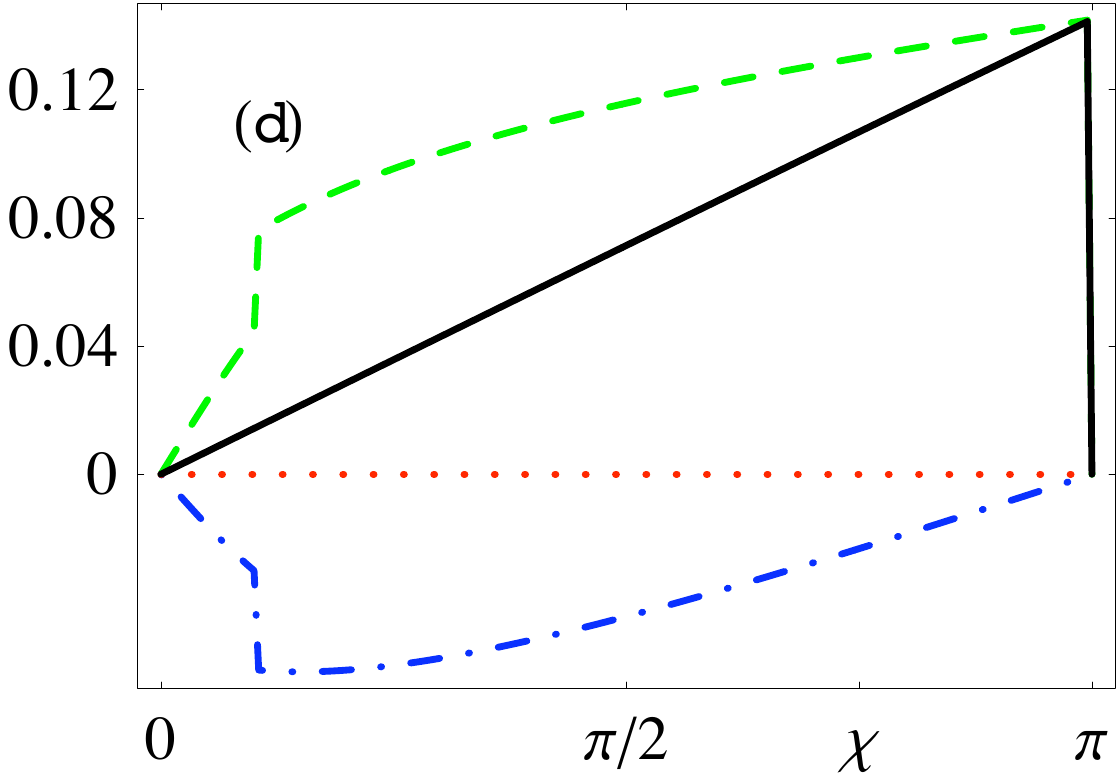}\\
\includegraphics[height=2.9cm ]{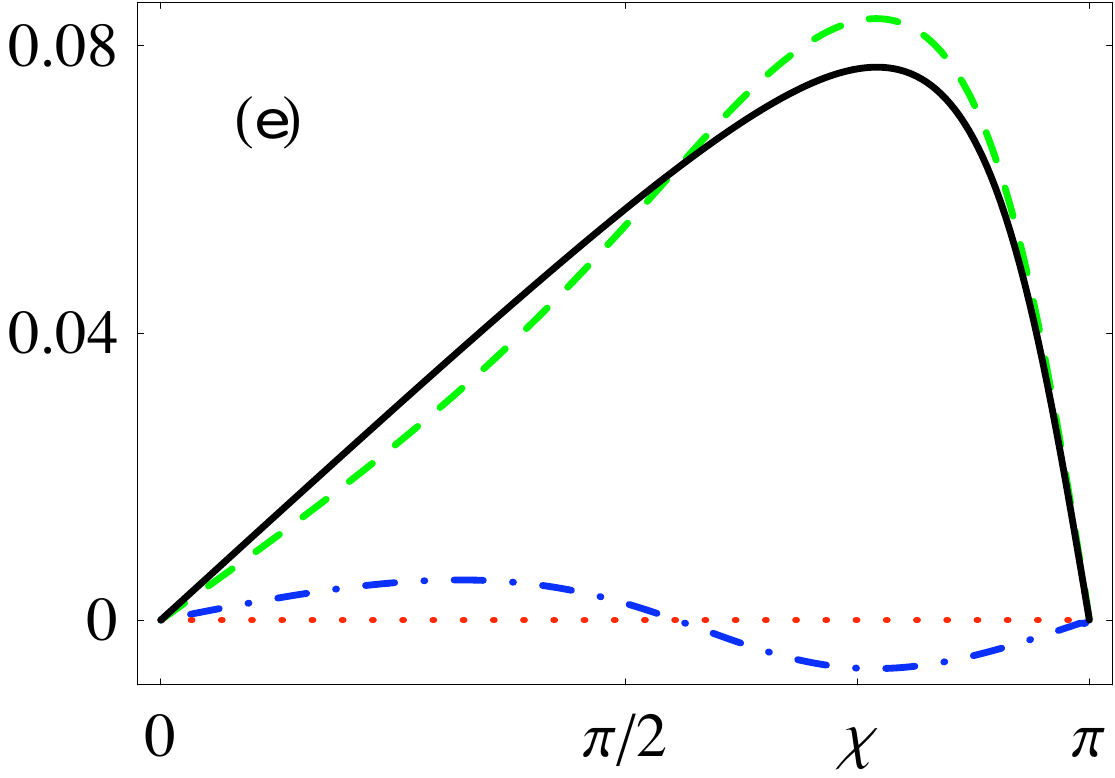}
\includegraphics[height=2.9cm ]{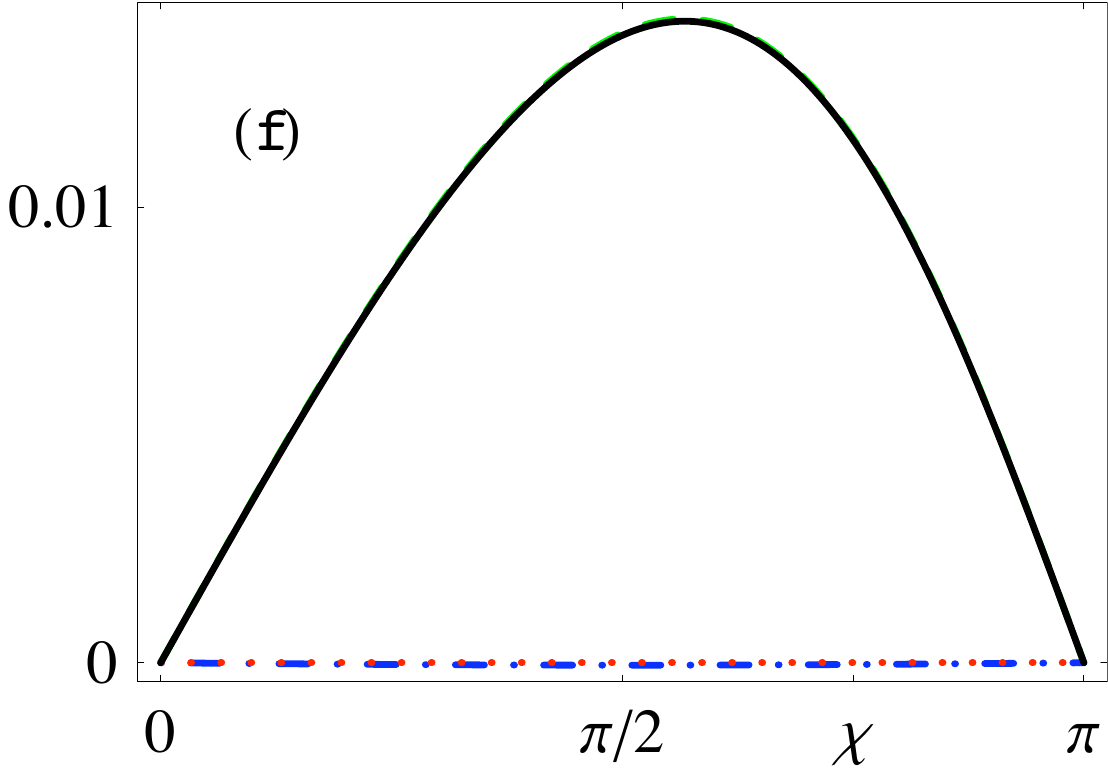}
\caption{Total Josephson current $J$ (solid curve),
$J_{\rm{cont}}$ (dotted-dashed curve), $J_{\rm{abs}}$ (dashed
curve) and $J_{\rm{nbs}}$ (dotted curve) as a function of
$\chi=\chi_{L}-\chi_{R}$ for a S-N-S junction with $M=8$,
$t_{S}=t_{T}=1$, $\Delta_{L}=\Delta_{R}=0.6$. The panels (a) to
(f) correspond to $t_{N}=1.5,1.2,1.0,0.744,0.6,0.3$. Energies are
in units of $|t_{S}|$.} \label{pannelli}
\end{figure}

\begin{figure}[h!]
\includegraphics[height=2.9cm ]{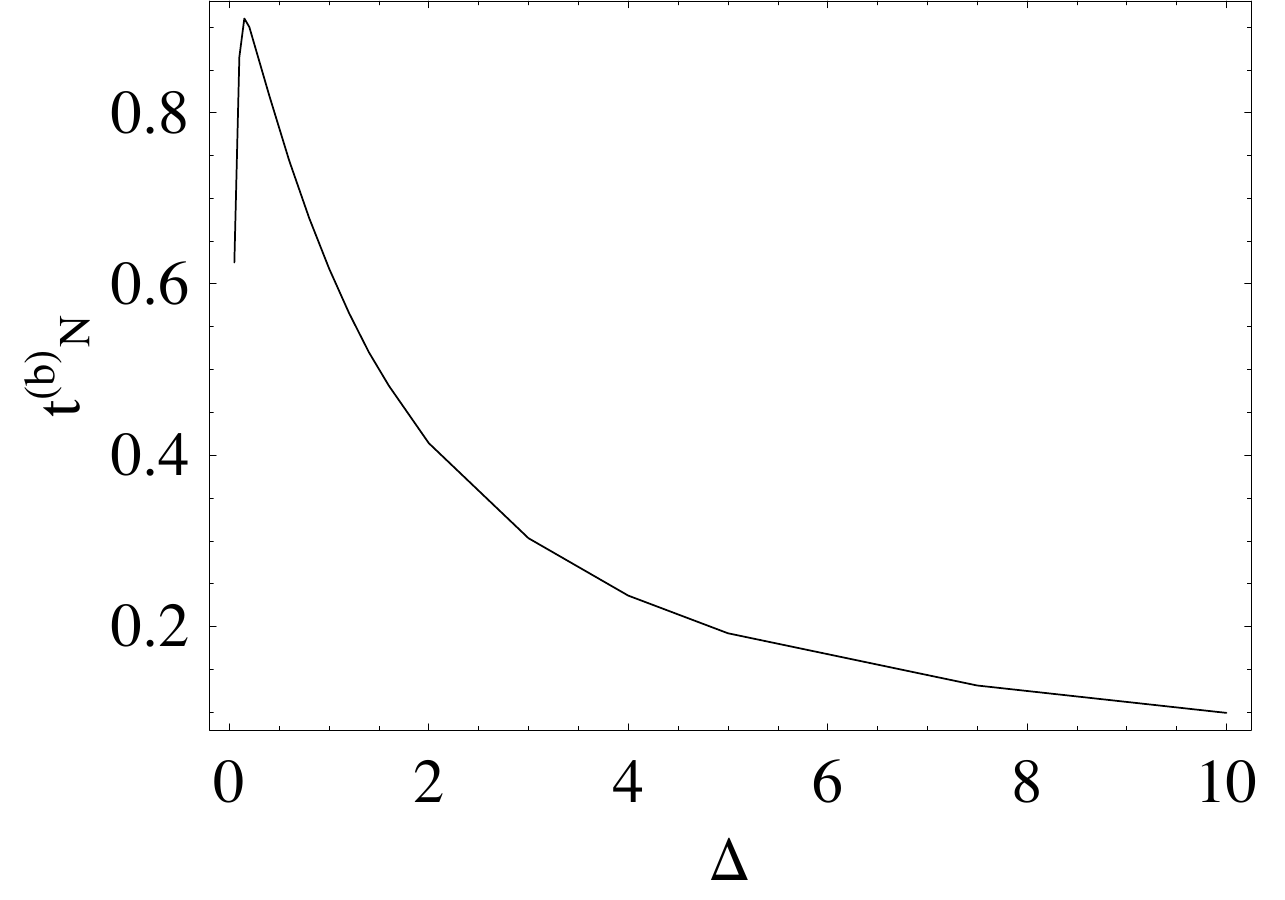}
\includegraphics[height=2.9cm ]{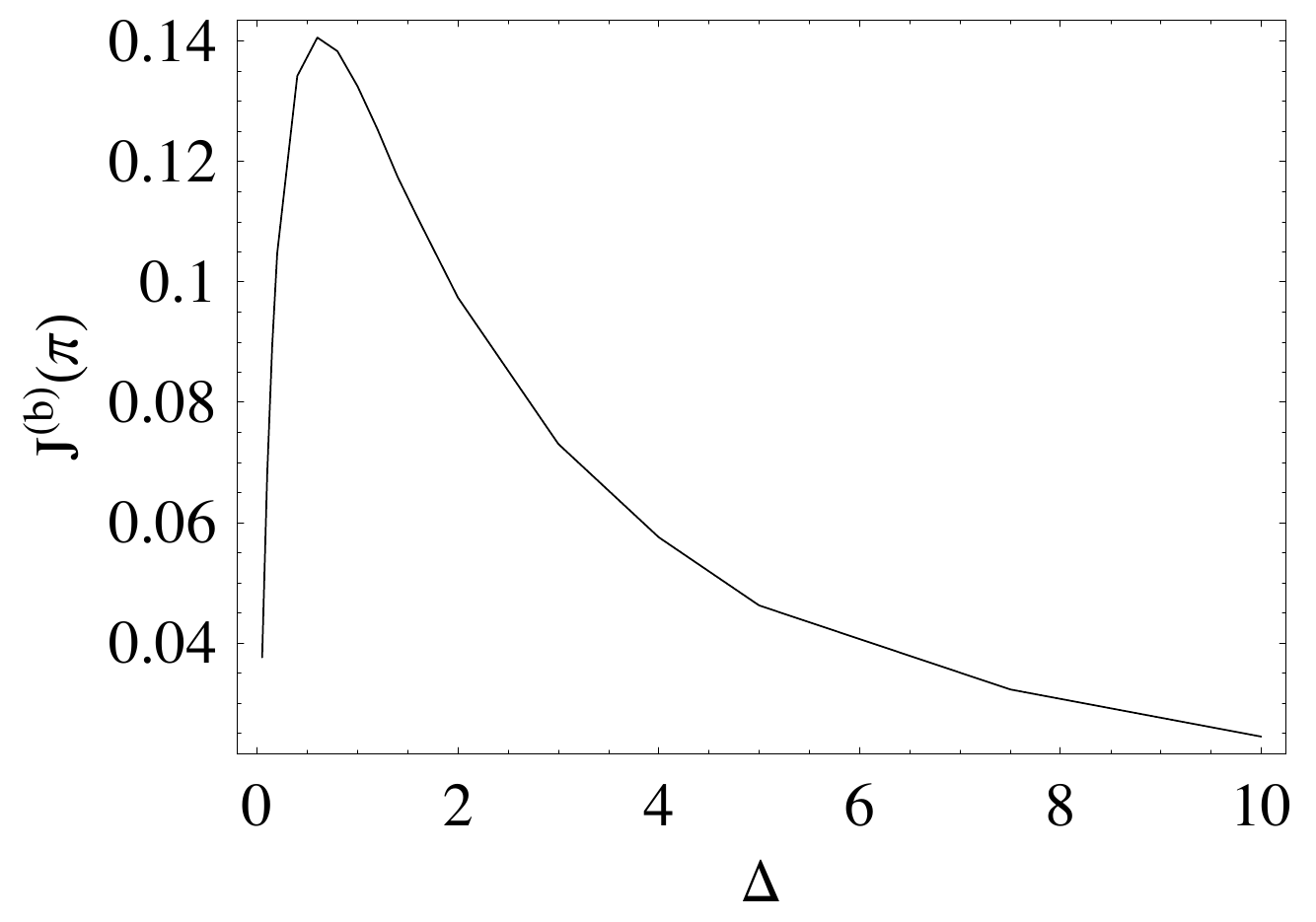}
\caption{ (Left panel) $t^{(b)}_{N}$ as a function of
$\Delta=\Delta_{L}=\Delta_{R}$. (Right panel) Critical current
$J(\pi)$ as a function of $\Delta$ calculated at
$t_{N}=t^{(b)}_{N}$. The rest of parameters are $M=8$,
$t_{S}=t_{T}=1$. Energies are in units of $|t_{S}|$.}
\label{tnbest}
\end{figure}

By following the approach described in the previous Section, we
specialize to the case of half-filled 1D leads as in
Fig.\ref{device} ($m=m^{\mathrm{1D}}$ and
$\tilde{\Delta}=\tilde{\Delta}^{\mathrm{1D}}$) and numerically
evaluate the dc Josephson current. In Fig.\ref{pannelli} we show
the current $J$ as a function of $\chi$ as well as the three
different contributions $J_{\rm{cont}},J_{\rm{abs}},J_{\rm{nbs}}$
for a chain of $M=8$ sites.
We notice that there is an optimum value of $t_{N}$ [see
Fig.\ref{pannelli} panel (d)] at which there is a non trivial
cancellation of the non-linear contributions $J_{\rm{cont}}$ and
$J_{\rm{abs}}$ and $J(\chi)$ becomes a straight line. In this
regime the Josephson  current is also maximized for every value of
$\chi$. We have further investigated this instance and found that
for any given $\Delta=\Delta_{L}=\Delta_{R}$, there exists an
optimum value of $t_{N} = t^{(b)}_{N}$ at which this property is
observed. In the left panel of Fig.(\ref{tnbest}) we plot
$t^{(b)}_{N}$ as a function of $\Delta$ for the same parameters as
in Fig.\ref{pannelli}. We have also observed that $t^{(b)}_{N}$ is
quite insensitive to the size $M$ of the normal region. In the
right panel of Fig.(\ref{tnbest}) we display the corresponding
critical current $J^{(b)}(\pi)$, i.e. the value of  the Josephson
current reached at $\chi=\pi$. The linearization of the
current-phase relation is also known as the Ishii's sawtooth
behavior\cite{ishii} and corresponds to perfect Andreev
reflection\cite{affleck2}.
We notice that the ABS contribution saturates the dc Josephson
current for $t_{N} =0.3$, see Fig.\ref{pannelli} panel (f). In the
next Section we consider the limit of long chains and identify a
regime for the occurrence of this saturation.

\begin{figure}[h!]
\includegraphics[height=3.9cm ]{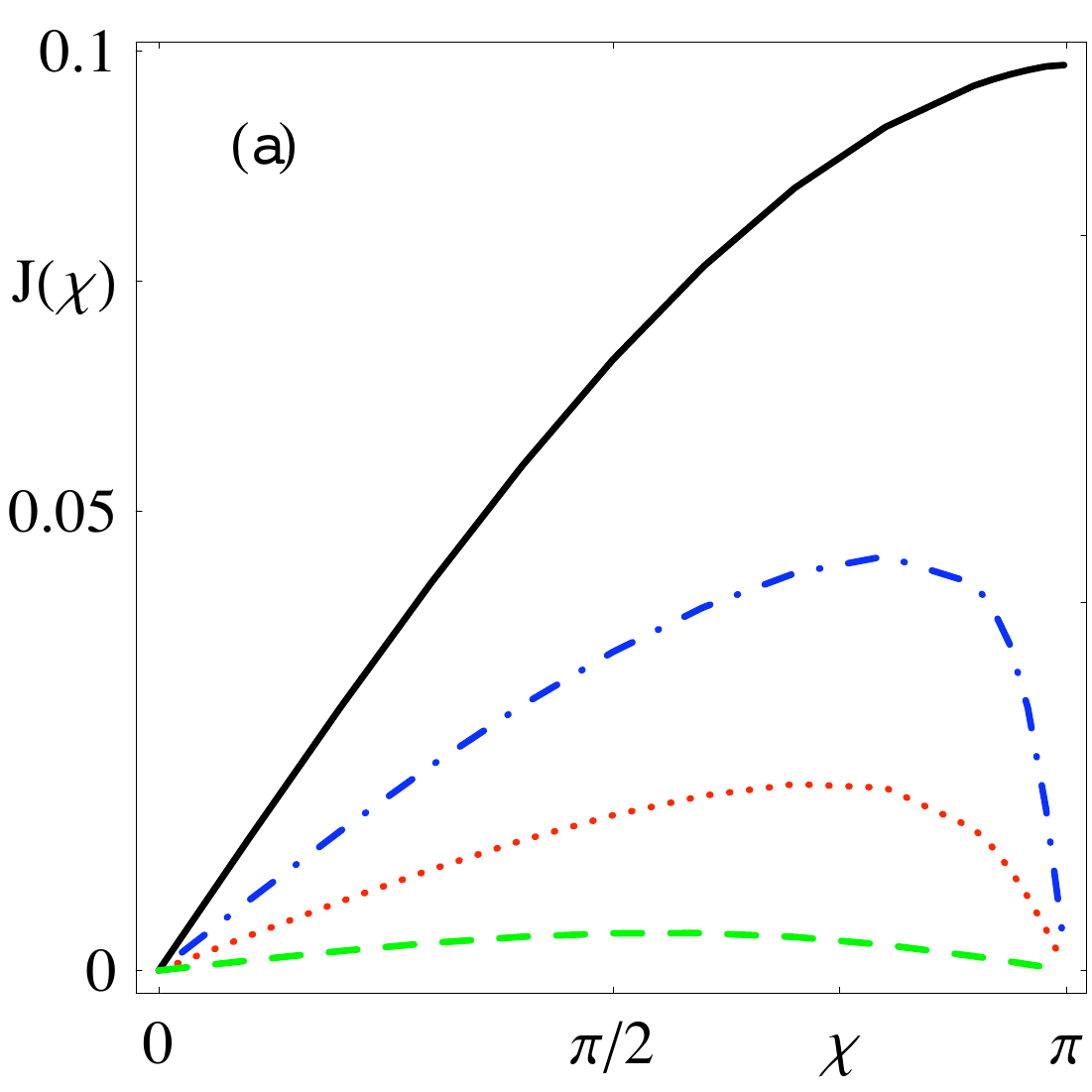}
\includegraphics[height=3.9cm ]{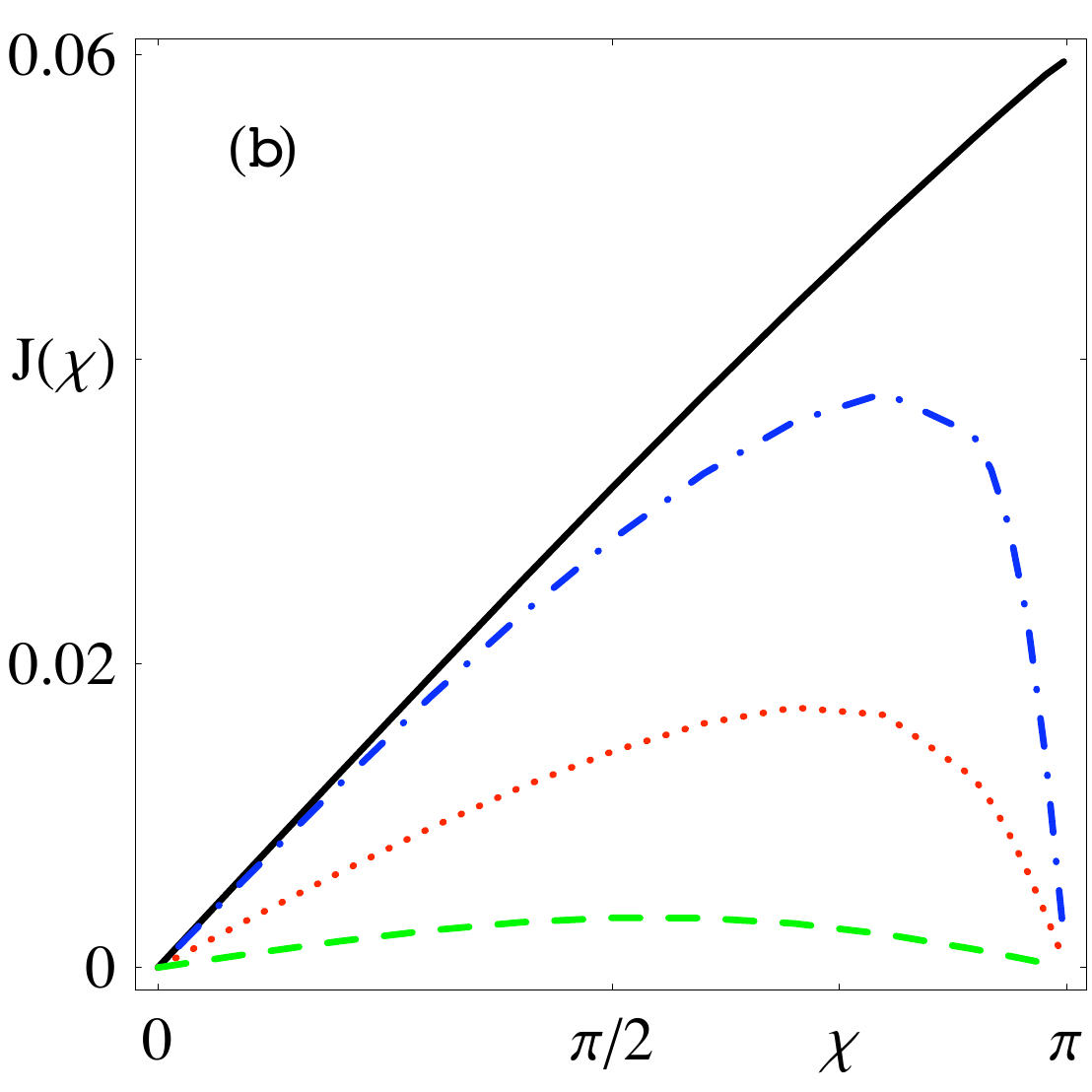}
\caption{Total Josephson current $J$ as a function of
$\chi=\chi_{L}-\chi_{R}$ for a short junction with $M=1$ (panel a)
and for a long junction with $M=51$ (panel b) for different values
of $\Delta_{L}=\Delta_{R}=\Delta=0.1,0.01,0.005,0.001$ (from top
to bottom). The rest of parameters are $t_{S}=1$, $t_{N}=3.8$,
$t_{T}=2$. For clarity, the curves corresponding to $\Delta=0.01$
and $\Delta=0.005$ have been multiplied by a factor 5, while
curves corresponding to $\Delta=0.001$ by a factor 10. Energies
are in units of $|t_{S}|$.} \label{cross}
\end{figure}

Finally we show that within our approach it is possible to
reproduce the crossover of the current-phase relation between
short and long S-N-S junctions, already discussed in previous
works\cite{kummel1}. In Fig.\ref{cross} we display $J(\chi)$ both
for a short junction ($M=1$) as well as for a long junction
($M=51$). It appears that for large superconducting order
parameter the current-phase relation evolves from a $\sin
(\chi/2)$-shaped curve\cite{beenakker} to a straight line by
passing from to $M=1$ to $M=51$. These results are in agreement
with the findings of Ref.\onlinecite{kummel1} where the change of
the order parameter $\Delta$ is due to a change of temperature.

\section{Limit of long normal region}
\label{seclong}

In this Section we study the Josephson current  in the limit of
long chains. By numerical inspection we have verified that
$J=J_{\rm{abs}}$ for $t_{N} \leq \Delta/2$, i.e. all the Josephson
current is carried by the ABS's. In this regime the number of
occupied ABS's equals exactly the number of sites $M$ of the
tight-binding chain. Thus the ABS's constitute a local basis set
with a good approximation. As a consequence no normal bound states
occur, while the amplitude of the current carrying continuum
states is exponentially suppressed in the normal region. The
current $J=J_{\rm{abs}}$ is obtained by calculating the
contribution $E^{\rm{tot}}_{\rm{abs}}$ of the ABS's to the total
energy
\begin{equation}
J(\chi)=2\frac{d E^{\rm{tot}}_{\rm{abs}}(\chi)}{d\chi} \,.
\label{occ1}
\end{equation}
To calculate the energy $E_{\rm{abs}}^{(k)}=2t_{N}\cos k$ of a
single ABS it is convenient to write
\begin{equation}
k=\frac{\pi j}{M+1} +\frac{\delta_{k}}{M+1} \quad , \quad j=1,...
\,M \,,
\label{kdelta}
\end{equation}
where $\delta_{k}$ is a $k$-dependent phase-shift. Following
Refs.\onlinecite {affleck1,affleck2} the total ABS energy can be
expressed as
\begin{eqnarray}
E^{\rm{tot}}_{\rm{abs}} &=&2(M+1)\int_{0}^{\pi}\frac{dk}{\pi}
E_{\rm{abs}}^{(k)} f(E_{\rm{abs}}^{(k)}) \nonumber
\\
&+&\int_{0}^{\pi}\frac{d k}{\pi} \left| \frac{d
E_{\rm{abs}}^{(k)}}{d k}\right| [\delta_{k,+}+
\delta_{k,-}] f(E_{\rm{abs}}^{(k)}) \nonumber \\
&+&\frac{1}{2} \frac{\pi
v_{F}}{M+1}\left[\left(\frac{\delta_{k_{F},+}}{\pi}
\right)^2+\left(\frac{\delta_{k_{F},-}}{\pi} \right)^2-\frac{1}{6}
\right] ,
\label{fumi}
\end{eqnarray}
where $\delta_{k,\pm}$ correspond to the two branches of ABS's,
$v_{F}=2t_{N}\sin k_{F}$ is the Fermi velocity and $k_{F}$ is the
Fermi momentum. For large $M$ the momentum $k$ is a
\textit{continuous} variable in the range $(0,\pi)$ and the
phase-shifts $\delta_{k}$ can be determined by inserting
Eq.(\ref{kdelta}) in Eq.(\ref{andreevt}) and expanding in powers
of $1/M$.  To lowest order Eq.(\ref{andreevt}) reduces to
\begin{equation}
\left( a_{k} \sin \delta_{k} + b_{k}\cos \delta_{k} \right)^{2}
=c_{k}^{2} \, , \label{delta}
\end{equation}
where $a_{k}=t_{N}^{2} +
(m_{k}^{2}-\tilde{\Delta}_{k}^{2})\cos(2k)-2t_{N}m_{k} \cos k$,
$b_{k}=2t_{N}m_{k} \sin k -(m_{k}^{2}-\tilde{\Delta}_{k}^{2})\sin
(2k)$ and $c_{k}^{2}=2(t_{N}\tilde{\Delta}_{k}  \sin  k ^{2}
(1-\cos \chi)$.  The solutions of Eq.(\ref{delta}) read
\begin{equation}
\delta_{k, \pm}= -\arctan \frac{b_{k}}{a_{k}} \pm
\frac{1}{2}\arccos \left(1-\frac{2c_{k}^{2}}{a_{k}^{2}+b_{k}^{2}}
\right) \, . \label{deltasol}
\end{equation}
Eq.(\ref{deltasol}) provides a generalization to nonvanishing
on-site potential of the phase-shifts found by Affleck et
al.\cite{affleck2}. Inserting Eq.(\ref{deltasol}) in
Eq.(\ref{fumi}) the Josephson current is obtained from
Eq.(\ref{occ1}). We notice that the combination $\delta_{k,+}+
\delta_{k,-}$ is independent of the phase difference $\chi$ for
any $\tilde{\Delta}_{k}$ and $m_{k}$. Therefore the dc Josephson
current reads
\begin{eqnarray}
J(\chi)&=& \frac{\pi v_{F}}{M+1}\frac{d}{d\chi}
\left[\left(\frac{\delta_{k_{F},+}}{\pi}
\right)^2+\left(\frac{\delta_{k_{F},-}}{\pi} \right)^2\right] \,.
\label{occ}
\end{eqnarray}
Below we specialize the analysis to 1D leads at half-filling
($k_{F}=\pi/2$). In this case $m_{k}^{\rm 1D}=0$, see
Eq.(\ref{embedding1}), and one can show that
\begin{equation}
\delta_{k_{F},\pm}=\pm \frac{1}{2}\arccos \left[
\frac{(t_{N}^{2}+\tilde{\Delta}_{k_{F}}^{2})^{2}+4t_{N}^{2}\tilde{\Delta}_{k_{F}}^{2}
(\cos \chi-1)}{(t_{N}^{2}+\tilde{\Delta}_{k_{F}}^{2})^{2}}
\right], \label{deltapm}
\end{equation}
with $\tilde{\Delta}_{k_{F}} = \tilde{\Delta}_{k_{F}}^{1D}$.
 We would like to stress that Eq.(\ref{deltapm}) has been
obtained starting from a microscopic model Hamiltonian, i.e.
without resorting to phenomenological effective on-site and
pairing potentials. The relation between the  effective pairing
potential $\tilde{\Delta}_{k_{F}}$ and the microscopic order
parameter $\Delta$ in Eq. (\ref{embedding2}) allows us to discuss
some relevant limiting cases in terms of physical quantities.

\begin{figure}[h]
\includegraphics[height=5cm ]{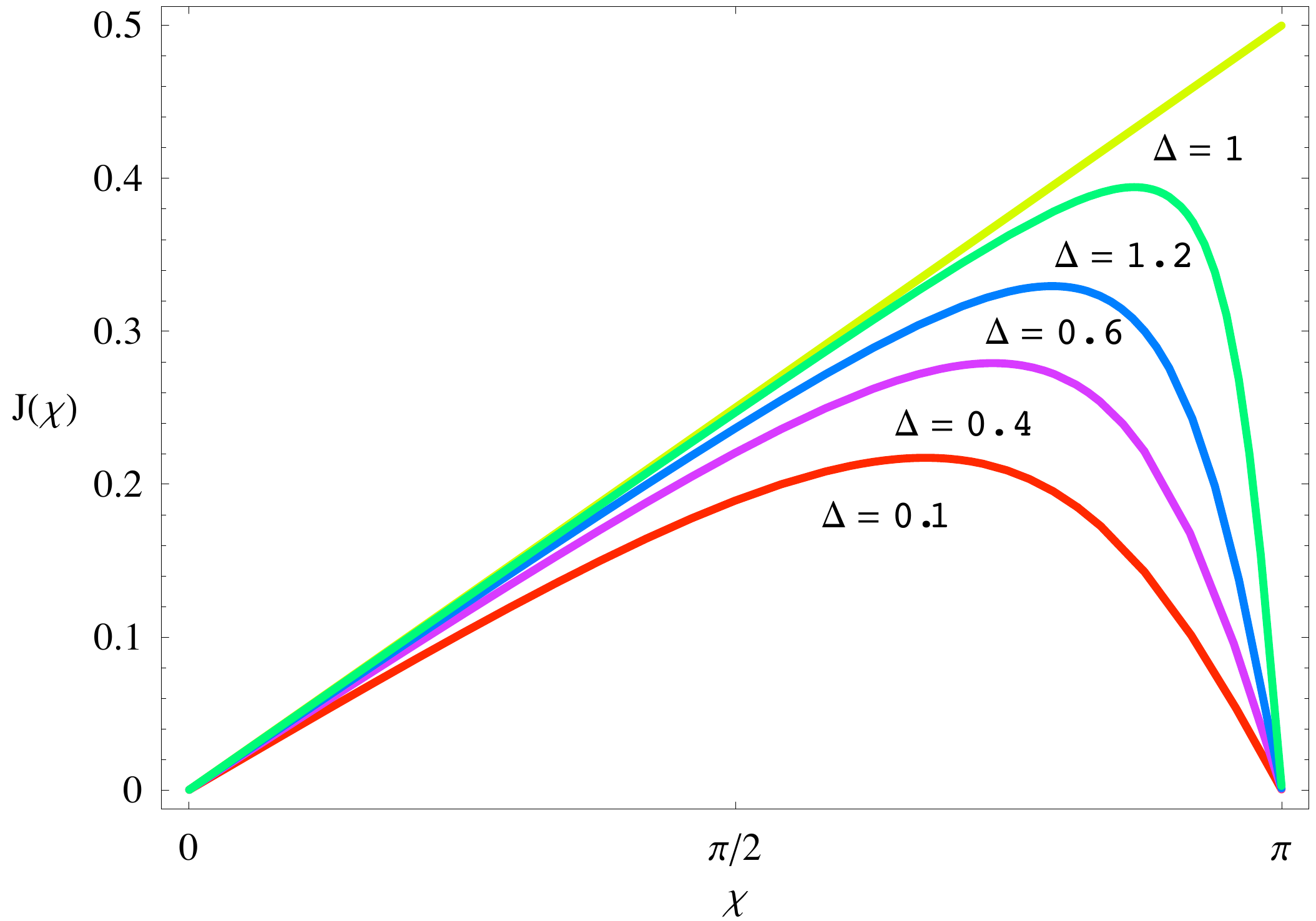}
\caption{ Josephson current as in Eq.(\ref{occ}) for different
values of $\Delta$ for $t_{N}=0.618$, $t_{S}=t_{T}=1$. For
$\Delta=1$ Eq.(\ref{perfect1}) is fulfilled. $J(\chi)$ is in unit
of $v_{F}/(M+1)$. Energies are in units of $|t_{S}|$.} \label{one}
\end{figure}

In Fig.(\ref{one}) we plot the Josephson current in Eq.(\ref{occ})
using for the phase-shift the result in Eq.(\ref{deltapm}). We fix
the values of the hopping parameters to be $t_{N}=0.618$,
$t_{S}=t_{T}=1$ and study how the current-phase relation depends
on $\Delta$. We notice that for $\Delta=1$ the current is linear
in the ranges $[0,\pi)$ and $(\pi,2\pi]$, with a sharp
discontinuity at $\chi=\pi$. This is the Ishii sawtooth
behavior\cite{ishii} already mentioned in the previous Section. In
that case, however, the sawtooth behavior was the result of a
perfect cancellation between the contribution of the continuous
states and of the ABS's. We also verified that the Josephson
current calculated by means of the brute-force numerical
evaluation of Eq.(\ref{joscurr}) at $t_{N} \leq \Delta/2$  is in
excellent agreement with the current evaluated as in
Eq.(\ref{occ}) already for $M \gtrsim 10 $.

As shown in Ref.\onlinecite{affleck2} the linear behavior of $J$
is due to perfect Andreev reflections which occur for
$t_{N}=\tilde{\Delta}^{\rm 1D}_{k_{F}}$, i.e.,
\begin{equation}
t_N=\frac{t_{T}^{2}}{2t_{S}^{2}}(\sqrt{4t_S^2+\Delta^2}-\Delta)
\,.
\label{perfect1}
\end{equation}
We recall that the above current corresponds to the total
Josephson current only for $t_{N}<\Delta/2$, which, together with
Eq.(\ref{perfect1}), implies
\begin{equation}
t_{N} \leq \frac{t_{T}^{2}}{\sqrt{2t_{T}^{2}+t_{S}^{2}}} \,.
\label{perfect2}
\end{equation}
Equations (\ref{perfect1},\ref{perfect2}) establish a regime in
which the Josephson current is entirely carried by the Andreev
bound states via perfect Andreev reflections.

Before concluding this Section we would like to observe that in
the WBL approximation the condition for perfect Andreev reflection
implies $t_{N}=\tilde{\Delta}^{\rm WBL}_{k_{F}}=\Gamma/2$, which
does not depend on the order parameter $\Delta$. The same
limitation of the WBL approximation emerges in the calculation of
the phase-shifts, see Eq.(\ref{deltapm}). Therefore the use of WBL
self-energies in superconducting transport through long normal
chains does not allow to study the dependence of the current-phase
relation on the physical order parameter.

\section{ac Josephson current}
\label{tdsec}

In this Section we consider the time-dependent current flowing
through the S-N-S junction after the switch-on of a dc bias
voltage. In order to get a sensible transient regime, we adopt the
so-called partition-free approach, in which the S-N-S system is
assumed to be contacted and in equilibrium before the external
bias is switched on\cite{cini,stefanucci}. The numerical results
contained in this Section are obtained by computing the exact
time-evolution of the system described in Eq.(\ref{ham}) with
finite 1D superconducting leads of length $\Lambda$ (see
Fig.\ref{device}). Without loss of generality we switch on the
bias at $t=0$. The biased Hamiltonian at positive times reads
\begin{equation}
\hat{H}(t)=\hat{H}_{N}+\hat{H}_{L}(t)+\hat{H}_{R}(t)+
\hat{H}_{T}-\mu(\hat{N}_{\uparrow}-\hat{N}_{\downarrow}) \,,
\end{equation}
where
\begin{eqnarray}
\hat{H}_{\alpha}(t)&=&\sum_{q} \left[(\varepsilon_{q}+U_{\alpha})
(c^{\dagger}_{q\alpha \uparrow}c_{q \alpha
\uparrow}-\tilde{c}^{\dagger}_{q \alpha \downarrow}
\tilde{c}_{q \alpha \downarrow}) \right. \nonumber \\
 &+&   \Delta_{\alpha}e^{i(\chi_{\alpha}+2U_{\alpha}t)} c^{\dagger}_{q \alpha
\uparrow}\tilde{c}_{q \alpha \downarrow}  \nonumber \\
&+& \left. \Delta_{\alpha}e^{-i(\chi_{\alpha}+2U_{\alpha}t)}
\tilde{c}^{\dagger}_{q \alpha \downarrow}c_{q \alpha \uparrow}
\right] \,,
\label{hambias}
\end{eqnarray}
and $U_{\alpha}$ are the dc bias voltages applied to lead
$\alpha$. We denote with $\underline{\mathbf{H}}(0)$ the matrix
representing the equilibrium Hamiltonian $\hat{H}$ of
Eq.(\ref{ham}) projected over one-particle states and with
$\underline{\mathbf{H}}(t)$ the corresponding matrix representing
$\hat{H}(t)$ of Eq.(\ref{hambias}) for $t>0$. The generic element
of $\underline{\mathbf{H}}(t \geq 0)$ is a $2 \times 2$ matrix in
the Bogoliubov-deGennes space
\begin{equation}
[\underline{\mathbf{H}}(t)]_{m,n}= \left(\begin{array}{cc}
H_{m,n}(t)& \Delta_{m,n}(t)
\\\Delta_{m,n}^{\ast}(t)&-H_{m,n}(t)
\end{array}\right) ,
\end{equation}
where $m,n=1, ... , 2\Lambda+M$. According to the partition-free
approach, we first calculate the equilibrium configuration of the
contacted system by solving the secular problem
\begin{equation}
\sum_{n}[\underline{\mathbf{H}}(0)]_{m,n} \left(\begin{array}{c}
u_{k}(n)
\\v_{k}(n)
\end{array}\right) = E^{(k)} \left(\begin{array}{c}
u_{k}(m)
\\v_{k}(m)
\end{array}\right) ,
\end{equation}
and construct the initial lesser Green's function
\begin{eqnarray}
&&[\underline{\mathbf{G}}^{<}(0,0)]_{m,n}=i[f(\underline{\mathbf{H}}(0))]_{m,n}
\nonumber \\
&&= \sum_{k} i f(E^{(k)})  \left(\begin{array}{cc} u_{k}^{\ast}(m)
u_{k}(n)& u_{k}^{\ast}(m) v_{k}(n)
\\v_{k}^{\ast}(m)
u_{k}(n)&v_{k}^{\ast}(m) v_{k}(n)
\end{array}\right).
\label{gless0}
\end{eqnarray}
The initial states are then propagated in time according to the
time-dependent Bogoliubov-deGennes equations
\begin{eqnarray}
i\frac{d}{dt}u_{k}(m,t)&=&\sum_{n}\left[ H_{m,n}(t)
u_{k}(n,t)+\Delta_{m,n}(t)v_{k}(n,t) \right] \nonumber\\
i\frac{d}{dt}v_{k}(m,t)&=&\sum_{n}\left[- H_{m,n}(t)
v_{k}(n,t)+\Delta_{m,n}(t)u _{k}(n,t) \right], \nonumber \\
\end{eqnarray}
which are solved by
\begin{equation}
\left(\begin{array}{c} u_{k}(m,t)
\\v_{k}(m,t)
\end{array}\right)=\sum_{n}\left[ Te^{-i \int_{0}^{t} d\tau
\underline{\mathbf{H}}(\tau)} \right]_{m,n} \left(\begin{array}{c}
u_{k}(n,0)
\\v_{k}(n,0)
\end{array}\right),
\label{uv}
\end{equation}
with initial condition  $u_{k}(m,0)=u_{k}(m)$ and
$v_{k}(m,0)=v_{k}(m)$ and $T$ the time-ordering operator. The
lesser Green's function $\underline{\mathbf{G}}^{<}(t,t)$ has the
same form as the r.h.s of Eq.(\ref{gless0}) with $u_{k}(m)$ and
$v_{k}(m)$ replaced by $u_{k}(m,t)$ and $v_{k}(m,t)$. Expressing
the time-dependent wavefunctions as in Eq.(\ref{uv}) it is
straightforward to show that
\begin{equation}
\underline{\mathbf{G}}^{<}(t,t)=Te^{-i \int_{0}^{t} d\tau
\underline{\mathbf{H}}(\tau)} \,\, \underline{\mathbf{G}}^{<}(0,0)
\,\, Te^{i\int_{0}^{t} d\tau \underline{\mathbf{H}}(\tau) } \, ,
\label{gless}
\end{equation}
We notice from Eq.(\ref{hambias}) that $\hat{H} (t)$ has an
explicit time-dependence (the time-dependent phase of the order
parameter) and hence the evolution operator is not the exponential
of a matrix albeit the bias is constant in time. This problem is
solved by discretizing the time and calculating the evolution of
the lesser Green's function within a time-stepping procedure
\begin{equation}
\underline{\mathbf{G}}^{<}(t_{j},t_{j}) \approx   e^{-i
\underline{\mathbf{H}}(t_{j})\delta t} \, \,
\underline{\mathbf{G}}^{<}(t_{j-1},t_{j-1}) \,\,
  e^{i
\underline{\mathbf{H}}(t_{j})\delta t} \, ,
\end{equation}
where $t_{j}=j\delta t$, $\delta t$ is a small time step and $j$ a
positive integer. The time dependent current at the left interface
is calculated from Eq.(\ref{current}).
The above approach allows us to reproduce the time evolution of
the infinite-leads system up to a time $T_{\mathrm{max}} \approx
2\Lambda/v$, where $v$ is the maximum velocity for an occupied
one-particle state. For $t \gtrsim T_{\mathrm{max}}$ high-velocity
particles have time to propagate till the far boundary of the
leads and back, yielding undesired finite-size effects in the
calculated current\cite{tdspin}. For this reason we set $\Lambda$
such that $2\Lambda/v$ is much larger than the time at which the
stationary oscillatory state is reached.

\begin{figure}[h!]
\includegraphics[height=5.cm ]{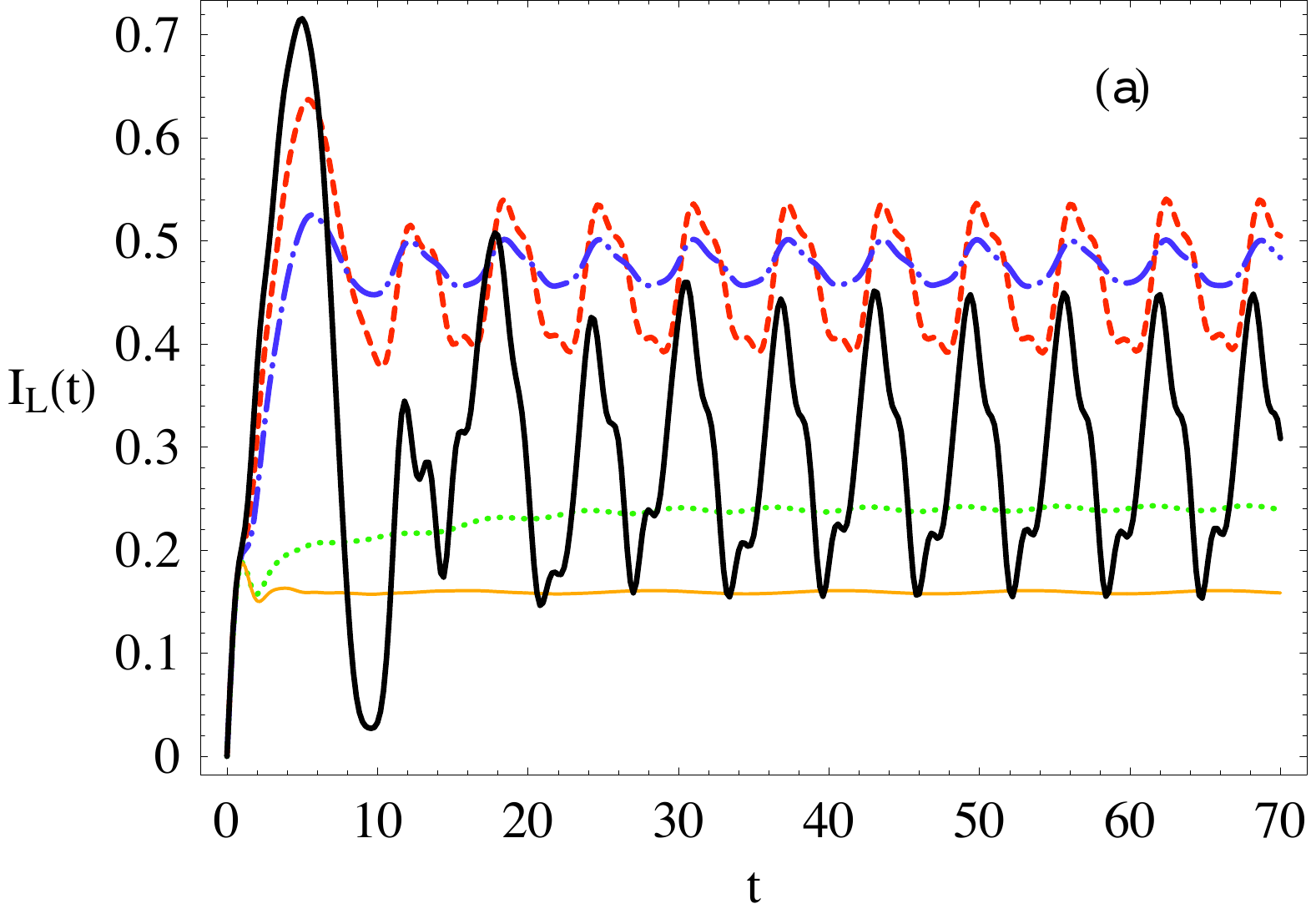}
\includegraphics[height=5.cm ]{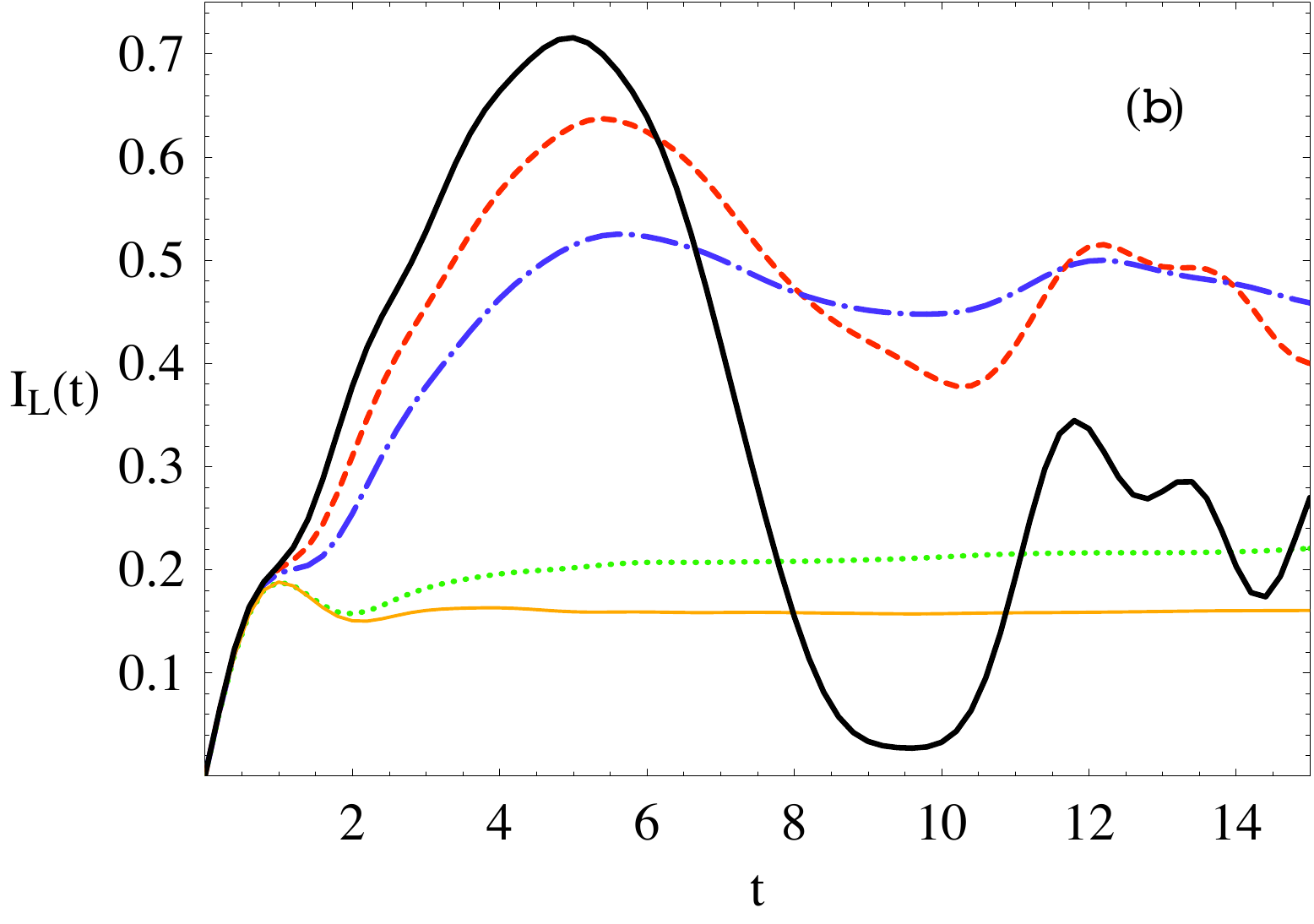}
\caption{Current $I_{L}(t)$ through the left interface for
different values of $\Delta_{L}=\Delta_{R}=\Delta=0$ (thin solid
curve), 0.1  (dotted curve), 0.5 (dotted-dashed curve), 0.7
(dashed curve), 1 (thick solid curve). The rest of parameters are
$M=1$, $\Lambda=80$, $\delta t=0.2$ $t_{S}=t_{T}=-1$,
$\chi_{L}=\chi_{R}=0$, $U_{L}=-U_{R}=0.25$. Panel (b) displays a
magnification of the transient regime for $0<t<15$. Energies are
in units of $|t_{S}|$, while time and $\delta t$ are in units of
$1/|t_{S}|$.} \label{tdjoseph}
\end{figure}

\begin{figure}[h!]
\includegraphics[height=5.cm ]{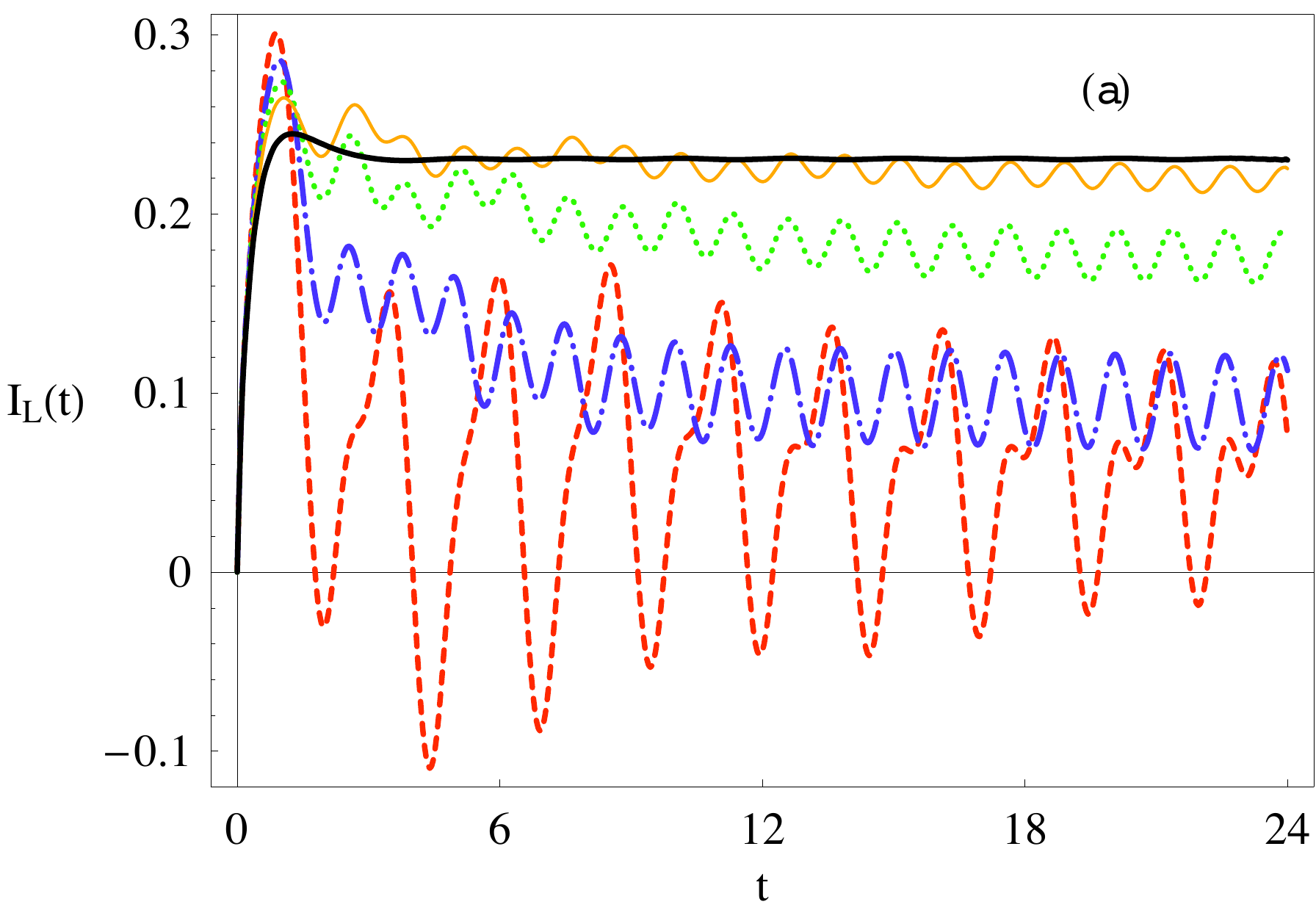}
\includegraphics[height=5.cm ]{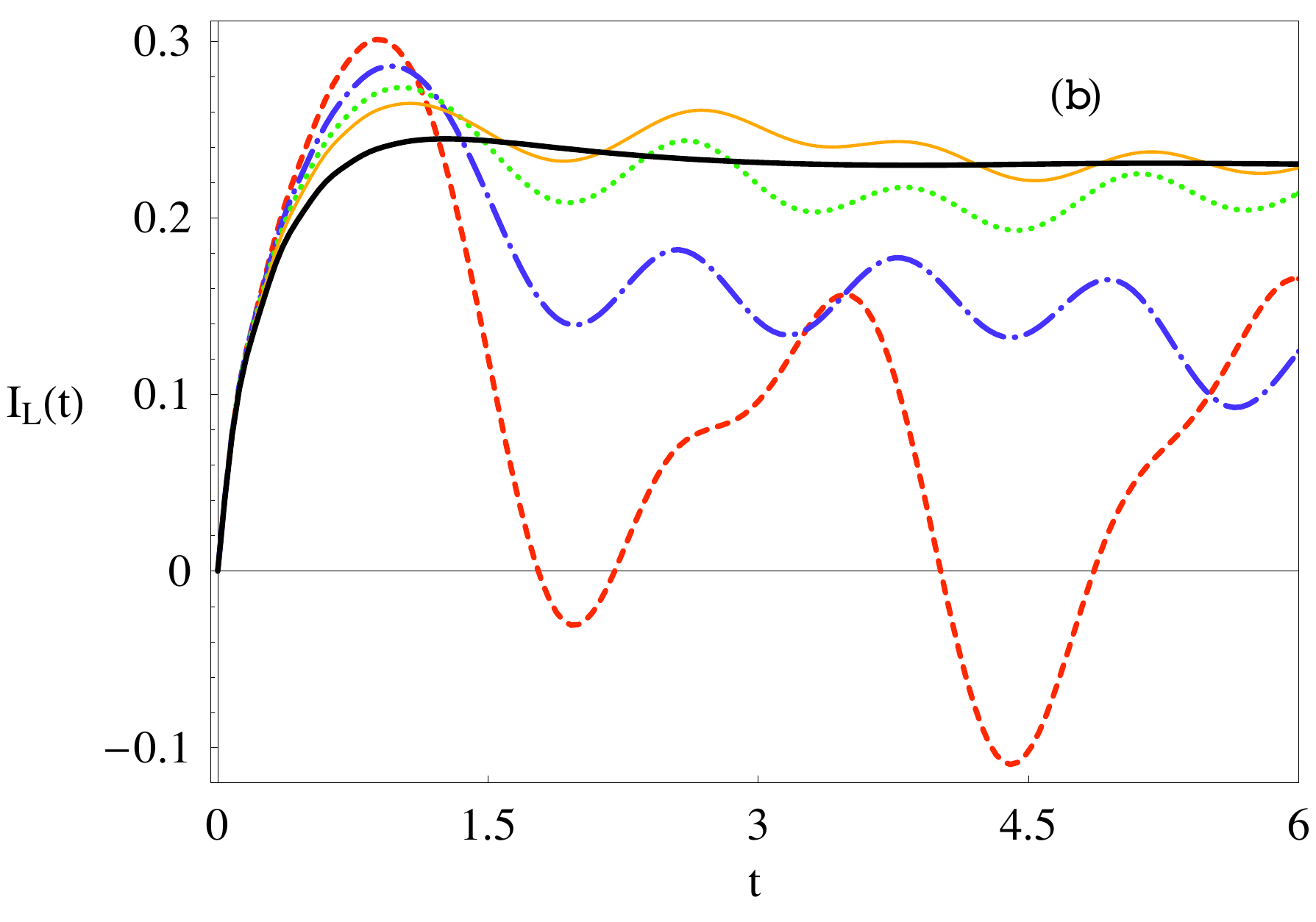}
\caption{Current $I_{L}(t)$ through the left interface for
different values of $\Delta_{L}=\Delta_{R}=\Delta=0$ (thick solid
curve), 0.25  (thin solid curve), 0.35 (dotted curve), 0.5
(dotted-dashed curve), 0.75 (dashed solid curve). The rest of
parameters are $M=1$, $\Lambda=6000$, $\delta t=0.1$ $t_{S}=100$,
$t_{T}=4.47$ (i.e. $\Gamma=2t_{T}^{2}/t_{S}=0.4$),
$\chi_{L}=\chi_{R}=0$, $U_{L}=-U_{R}=0.5$. Panel (b) displays a
magnification of the transient regime for $0<t<6$. The time
propagation has been obtained by retaining only the one-particle
states in Eq.(\ref{hambias}) with energy $-10 \leq\varepsilon_{q}
\leq 10$. We have checked that within this choice the results with
$\Delta=0$ perfectly agree with ones of
Ref.\onlinecite{stefanucci} obtained within the WBL approximation.
Energies are in units such that $\Gamma=0.4$, while time and
$\delta t$ are in units of $1/\Gamma$.} \label{tdwbl}
\end{figure}

In Fig.\ref{tdjoseph} we plot the time-dependent current through a
single-dot junction ($M=1$) for different values of the
superconducting order parameter $\Delta_{L}=\Delta_{R}=\Delta$,
ranging from 0 to 1. In panel (b) we display a magnification of
the transient regime.
It appears that the transient dynamics becomes slower as $\Delta$
is increased. This is due to the fact that at bias $U\approx
2\Delta/n$,  an incident electron coming from the left
superconducting lead undergoes about $n$ Andreev reflections
inside the central region before being transmitted to the right
lead. We also verified that, at fixed $\Delta$, the transient
timescale grows by reducing the bias voltage (not shown). A
qualitatively similar behavior is observed in Fig.\ref{tdwbl},
where the hopping in the superconducting leads is taken about two
orders of magnitude larger that all the other energy scales, in
the spirit of the WBL approximation. Another interesting observed
feature is that the dc component of the current $\bar{I}_{L}$ in
Fig.\ref{tdjoseph} displays a non-linear behavior with $\Delta$.
In particular $\bar{I}_{L}$ increases with $\Delta$ passing from 0
to 0.5, but decreases by further increasing $\Delta$ form 0.5 to
1. Such behavior, however, is not seen in Fig.\ref{tdwbl}, where
$\bar{I}_{L}$ is a monotonically decreasing function of $\Delta$.

\begin{figure}[h!]
\includegraphics[height=2.9cm ]{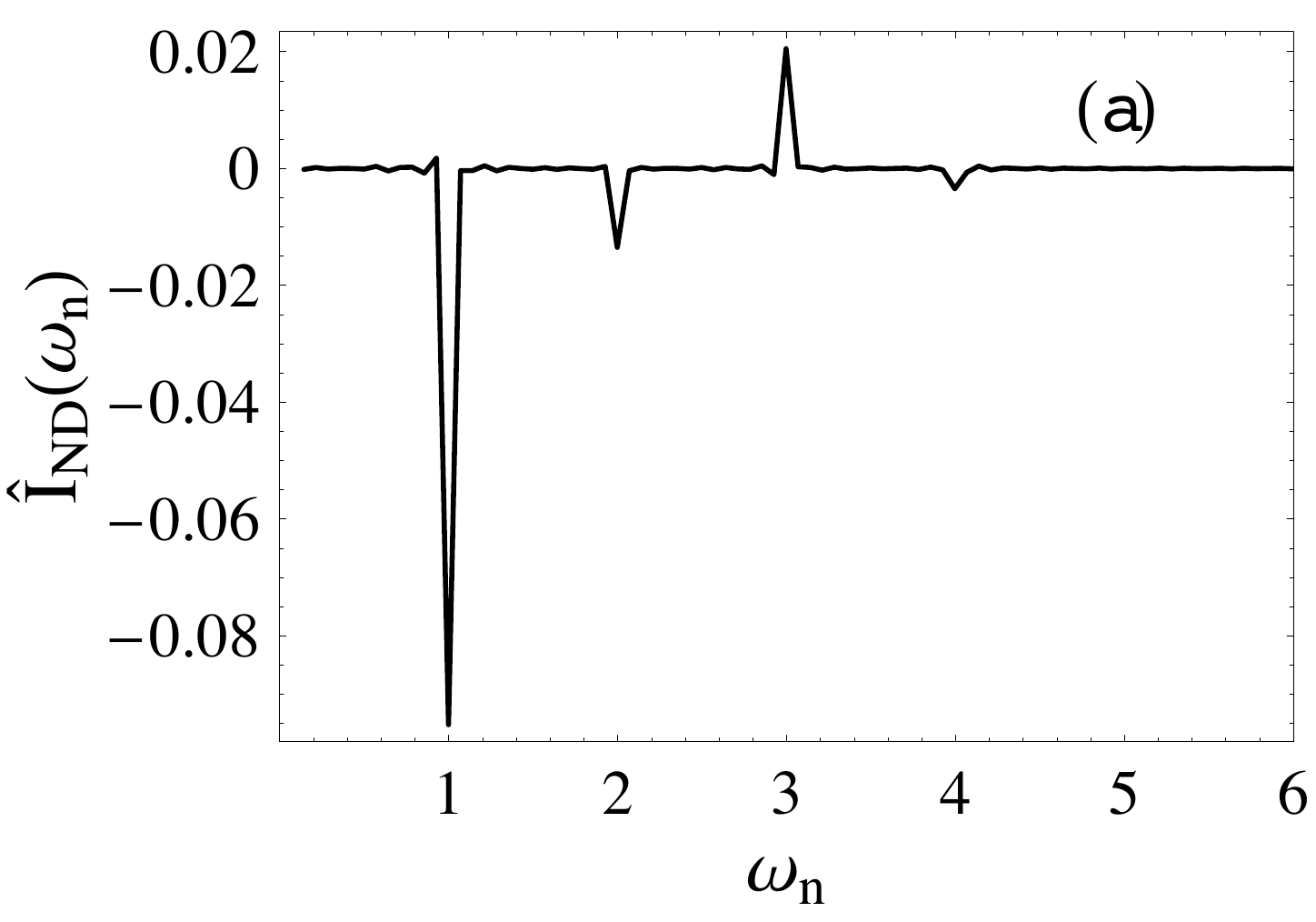}
\includegraphics[height=2.9cm ]{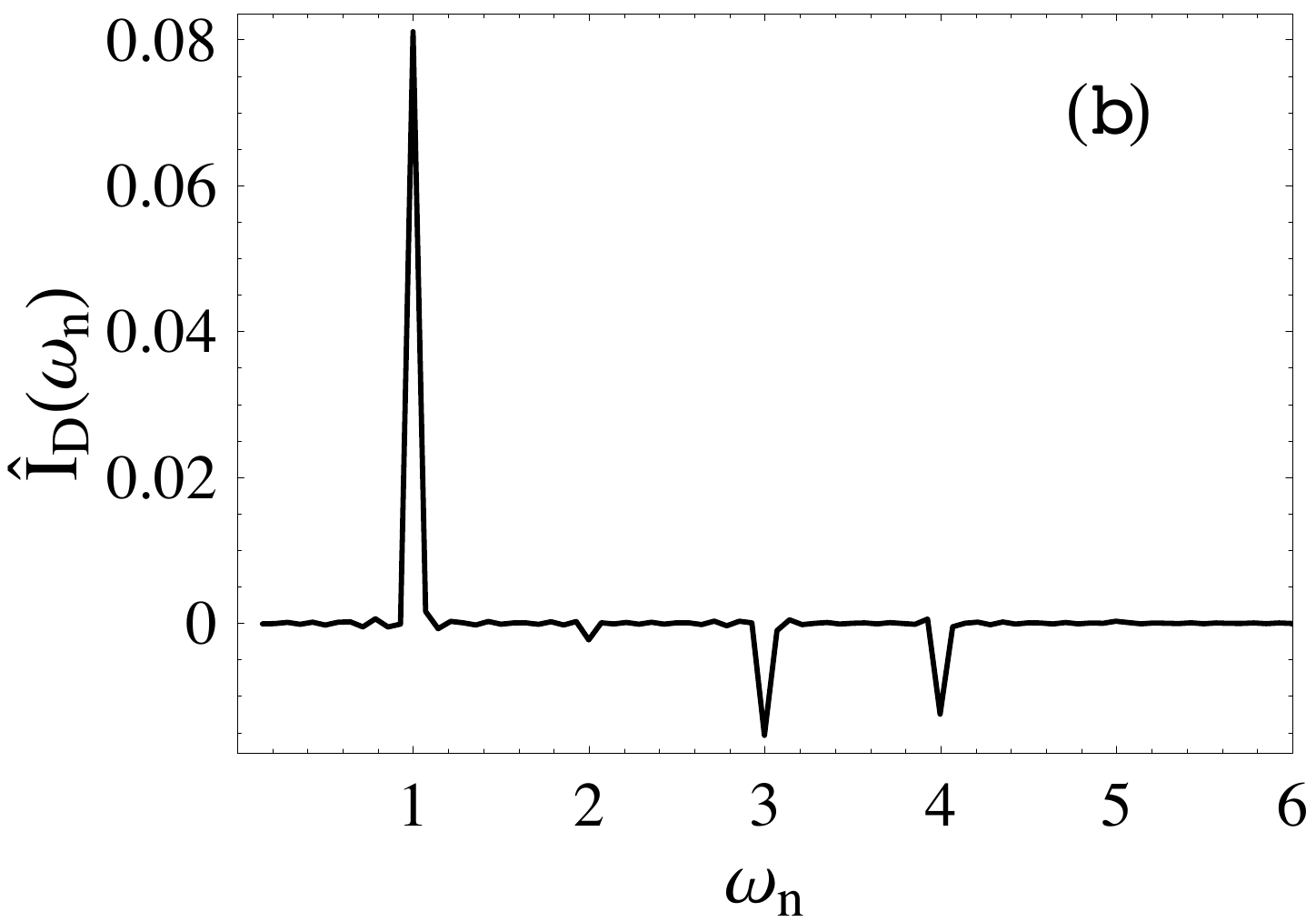}
\caption{Non-dissipative coefficient $\hat{I}_{\rm ND}$ (panel a)
and dissipative coefficient $\hat{I}_{\rm D}$ (panel b) obtained
from the discrete Fourier transform of $I_{L}(t)$ as described in
the main text. They are calculated using 2000 equidistant points
of $I_{L}(t)-\bar{I}_{L}$ with $t$ in the range $(50,140)$. In
this plot $\Lambda=150$, $\delta t =0.05$, $\Delta=1$ and the
Josephson frequency is $\omega_{J}=2(U_{L}-U_{R})=1$. The rest of
the parameters are the same as in Fig.\ref{tdjoseph}. Energies a
and frequency are in units of $|t_{S}|$.} \label{jomega}
\end{figure}

\begin{figure}[h!]
\includegraphics[height=5.cm ]{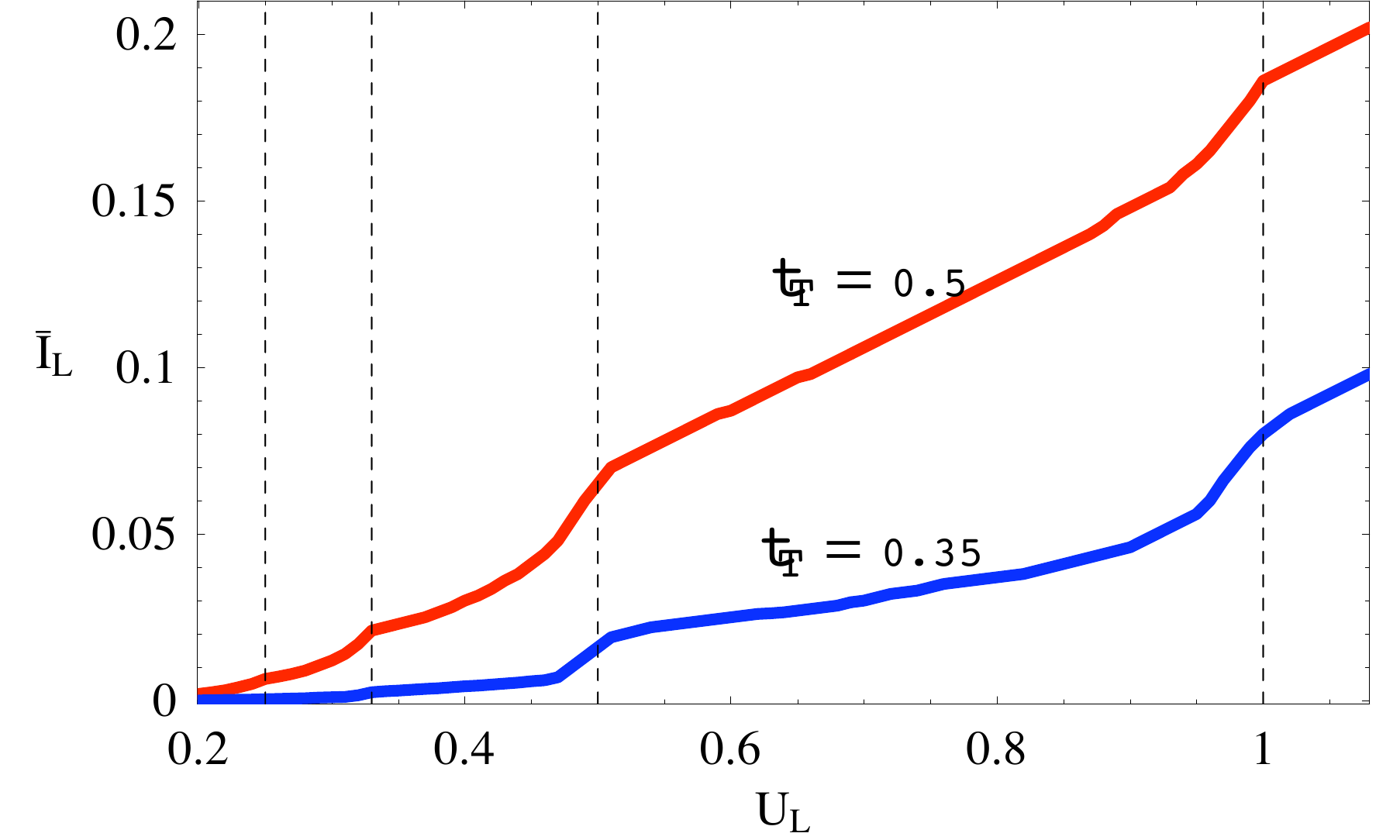}
\caption{Current-voltage characteristics ($\bar{I}_{L}$ vs
$U_{L}$) of the S-S junction for different values of the hopping
$t_{T}$. The rest of parameters are $U_{R}=0$, $t_{S}=-1$ and
$\Delta=0.5$. The vertical dotted lines denote the values
$U_{L}=2\Delta/n$ ($n$=1,2,3,4) at which multiple Andreev
reflections are expected. Energies are in units of $|t_{S}|$.}
\label{iv}
\end{figure}

At long time the current  $I_{L}(t)$ displays the well known ac
Josephson behavior, with persistent oscillations at multiple
frequencies of the fundamental Josephson frequency
$\omega_{J}=2(U_{L}-U_{R})$. To investigate the stationary
oscillations we performed a discrete Fourier transform of
$I_{L}(t)$ in the time window $(T_{\rm min},T_{\rm max})$ where
$T_{\rm min}$ is much larger than the transient timescale.
Denoting with $N_{f}$ the number of time steps in the time window,
the Fourier components of $I_{L}(t)$ are defined according
to\cite{stefanucci2,tdspin}
\begin{equation}
\hat{I}(\omega_{n})=\frac{1}{N_{f}} \sum_{j=1}^{N_{f}} e^{-i
\omega_{n}t_{j}} [I_{L}(t_{j})-\bar{I}_{L}] \, ,
\end{equation}
where $\omega_{n}=2\pi n /(N_{f}\delta t)$. In Fig.\ref{jomega} we
plot the dissipative contribution $\hat{I}_{\rm D}(\omega_{n})=2
{\rm Re} \hat{I}(\omega_{n}) $ and the nondissipative one
$\hat{I}_{\rm ND}(\omega_{n})=-2 {\rm Im} \hat{I}(\omega_{n}) $ to
the current\cite{cuevas1,guo}. The first four harmonics are
clearly visible and the fundamental component is the dominant one.
We also observe that the amplitude of the harmonics is not a
monotonically decreasing function of the frequency. The above
procedure provides an alternative method to perform the spectral
decomposition of the ac Josephson current. Our time-dependent
approach is not limited to dc biases and the same computational
effort is required to study ac or more complicated time-dependent
biases.
From our numerical time-dependent simulations, it is also possible
to extract the current-voltage characteristics of the junction. In
Fig.(\ref{iv}) we show $\bar{I}_{L}$ as a function of the applied
dc bias for a S-S junction ($M=0$).
The system consists of two 1D superconductors connected to each
other via a hopping integral $t_{T}$ between the boundary sites of
the $L$ and $R$ leads. We observe a well defined sub-gap structure
characterized by current kinks at $U_{L}-U_{R}=2\Delta/n$, a
feature already pointed out in previous works within the WBL
approximation\cite{cuevas1,cuevas2,guo}. We have also checked that
if the WBL is modelled with 1D leads (i.e. by taking $t_{S}\gg 1$
and $t_{T}=\sqrt{\Gamma t_{S}/2}$ with finite $\Gamma$), we
numerically recover the current-voltage characteristics already
obtained in previous works\cite{cuevas1,cuevas2}.

\begin{figure}[h!]
\includegraphics[height=5.cm ]{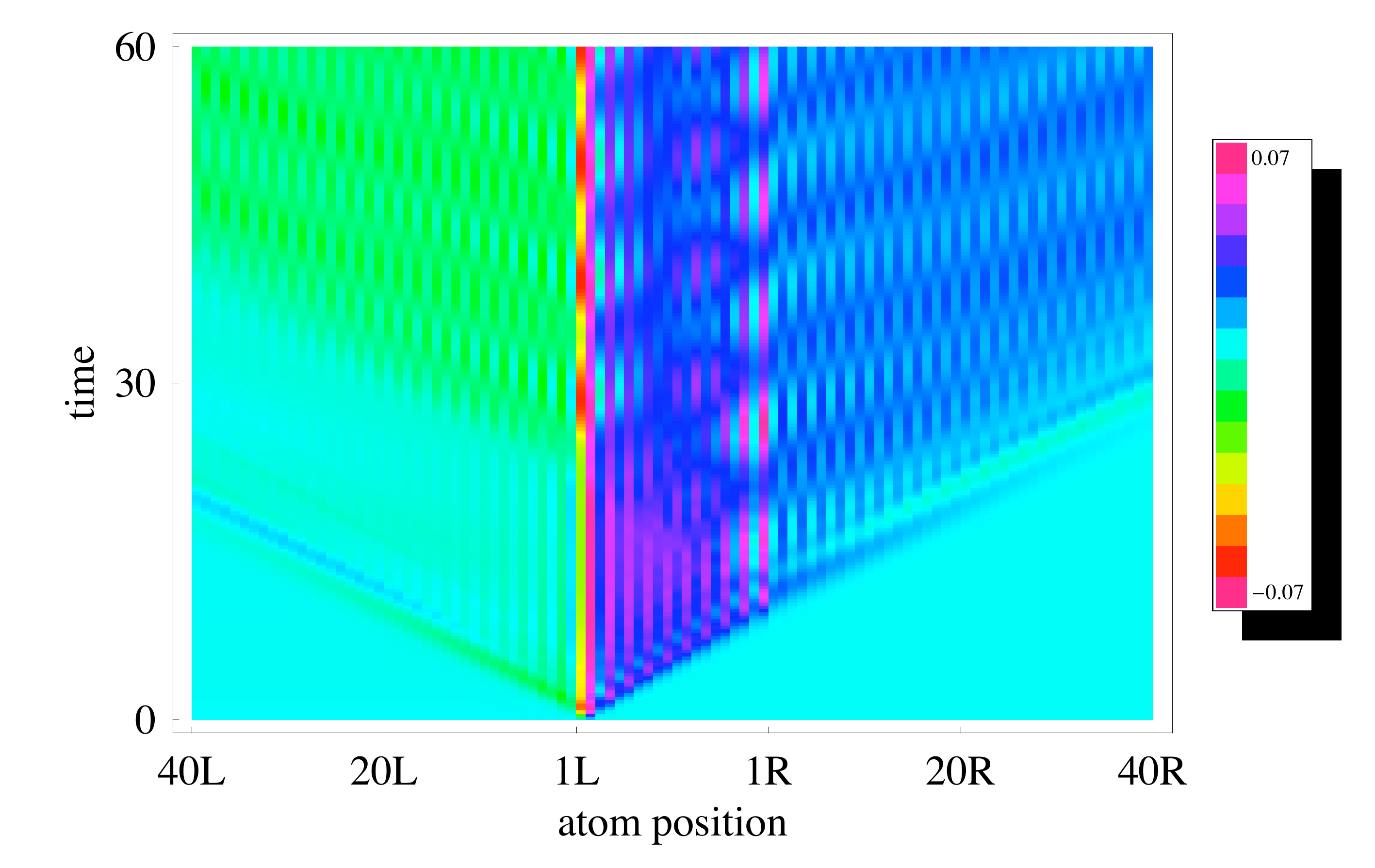}
\caption{Contour plot of the time-dependent variation of density
for spin-up electrons $\delta n_{m \uparrow}(t)=n(t)_{m
\uparrow}-n(0)_{m \uparrow}$ as a function of the atomic position
$m$ along the 1D S-N-S system ($x$ axis) and time ($y$ axis).
$\delta n_{i \uparrow}(t)$ is displayed for the first 40 sites in
both leads and inside the $M=20$ sites of the normal region. The
rest of parameters are $\Lambda=100$, $\delta t =0.3$,
$t_{S}=1.2$, $t_{N}=1$, $t_{T}=0.8$, $\Delta_{L}=\Delta_{R}=0.2$,
$\chi_{L}=\chi_{R}=0$, $U_{L}=0.3$, $U_{R}=0$. Energies are in
units of $|t_{N}|$, while time and $\delta t$ are in units of
$1/|t_{N}|$. } \label{pattern}
\end{figure}

\begin{figure}[h!]
\includegraphics[height=5.cm ]{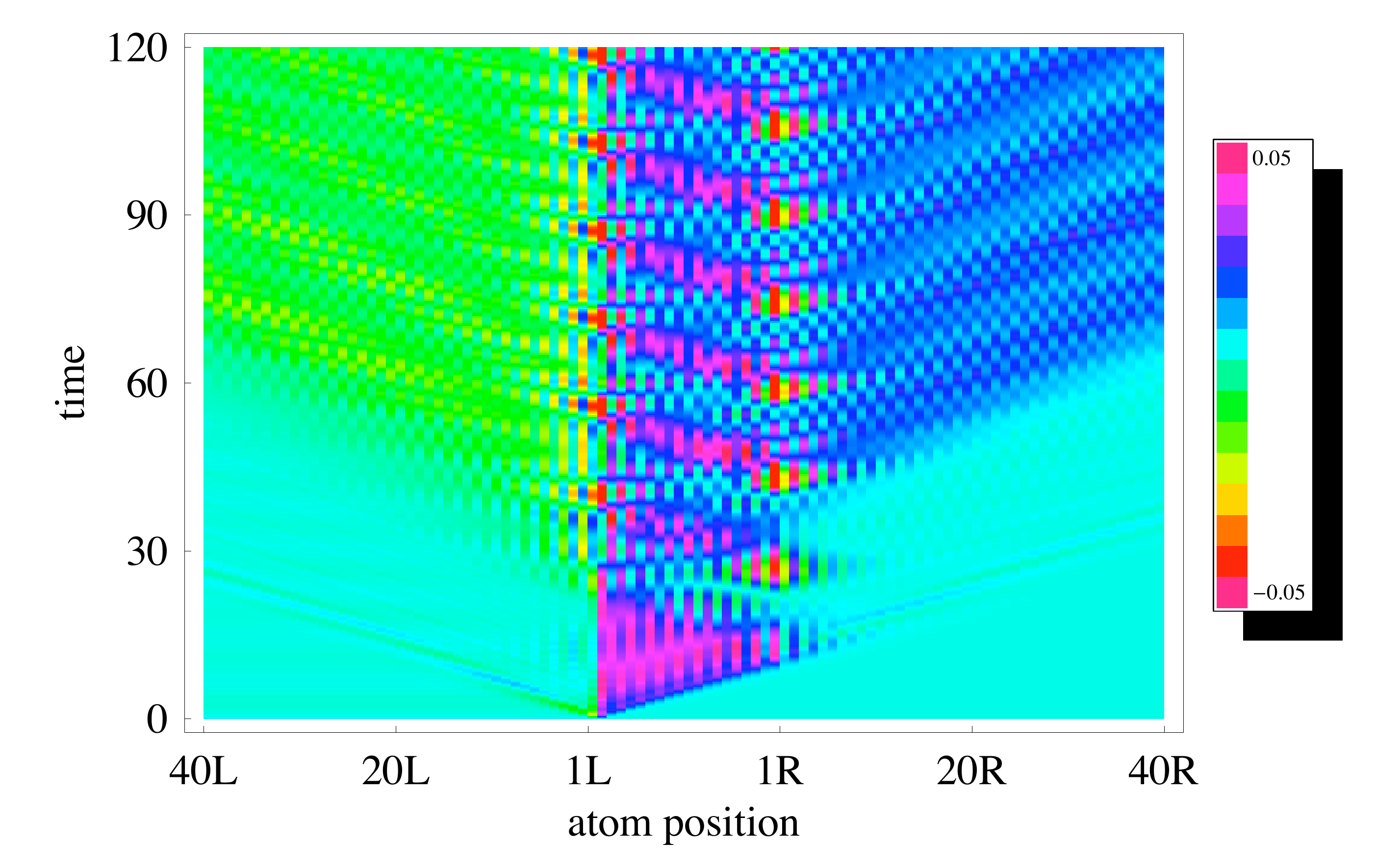}
\caption{Same as Fig.\ref{pattern}. The model parameters are:
$M=21$, $\Lambda=200$, $\delta t =0.3$, $t_{S}=1$, $t_{N}=1$,
$t_{T}=1.104$, $\Delta_{L}=\Delta_{R}=0.4$, $\chi_{L}=\chi_{R}=0$,
$U_{L}=0.2$, $U_{R}=0$. Energies are in units of $|t_{N}|$, while
time and $\delta t$ are in units of $1/|t_{N}|$ } \label{pattern2}
\end{figure}

Finally we have computed the time-dependent evolution of the
spin-up electron density according to
\begin{equation}
n_{m \uparrow}(t)=-i
\Big{(}[\underline{\mathbf{G}}^{<}(t,t)]_{m,m} \Big{)}_{1,1} \,,
\end{equation}
where $m$ denotes a site of the S-N-S system and the matrix
element $(\dots)_{1,1}$ is taken over the Nambu space. We stress
that our approach allows us to determine $n_{m \uparrow}(t)$ not
only in the normal region, but also inside the superconducting
leads\cite{vanleuw}. This is a clear advantage with respect to the
WBL approximation, in which only the dynamics of the normal region
can be described. In Fig.\ref{pattern} we show the density
variation $\delta n_{m \uparrow}(t)=n(t)_{m \uparrow}-n_{m
\uparrow}(0)$ as a function of the atomic position $m$ along the
1D S-N-S system and time. In this case a long junction with $M=20$
is considered. It is clearly seen at $t>0$ the perturbation
induced by the switch-on of the bias ($U_{L}\neq 0$ and $U_{R}=0$)
propagates both inside the $L$ lead (leftward) and the normal
region (rightward) with velocities $v_{S} \approx 2t_{S}$ and
$v_{N}\approx 2t_{N}$ respectively. At long time the density
displays stationary oscillations due to the stabilization of the
ac Josephson regime. In particular on the left lead the average
value of $\delta n_{m \uparrow}(t)$ is lower with respect to the
one in the right lead, since $U_{L}>U_{R}$. In Fig.\ref{pattern2}
we plot the transient behavior of the charge density for a
junction in which the (equilibrium) condition for perfect Andreev
reflection given in Eq.(\ref{perfect1}) has been imposed.
Remarkably we see that no appreciable density variation inside the
lead $R$ occurs before a dwelling time given by
\begin{equation}
t_{\rm{dwell}} \approx n \, t_{\rm{AR}} \, ,
\end{equation}
where $n=2\Delta/(U_{R}-U_{L})$ and where $t_{\rm{AR}}=M/v_{N}$ is
the time needed to cross the normal chain between two consecutive
reflections. Indeed for $t<t_{\rm{dwell}}$ an electron inside the
N region undergoes $n$ (almost) perfect Andreev reflections before
being transmitted through the right interface. The pattern of
these multiple reflections is clearly visible in
Fig.\ref{pattern2}, in which the model parameters are chosen in
order to have $n=4$ and $t_{\rm{dwell}} \approx 36$.

\section{Summary and conclusions}
\label{conclusions}

In this paper we have studied the dc and ac transport properties
of a tight-binding S-N-S junction. In the dc case we identified
three contributions to the dc Josephson current coming from the
Andreev bound states, normal bound states and continuum states.
The calculation of the latter contribution has been performed by
employing an exact embedding procedure which consists in
integrating out the superconducting degrees of freedom and in
expressing the Nambu-Gorkov Keldysh Green's function in terms of
the embedding self-energy. For the bound-state contributions we
calculated the phase derivative of the eigenenergies of all
occupied discrete states. The secular problem is cast in terms of
an effective energy-dependent Hamiltonian in which the on-site and
pairing potentials of the normal chain are renormalized via the
embedding self-energy. The bound-state eigenenergies of chains of
arbitrary length are determined from a general equation which
includes the full frequency dependence of the embedding
self-energy. The limit of long-chains allows for further analytic
manipulations and the ABS's contribution to the total dc Josephson
current is expressed in terms of energy-dependent phase-shifts.

For 1D superconducting leads we obtain an exact formula for the
embedding self-energy at half-filling. Explicit numerical results
have been presented for short and long chains, and different
regimes have been analyzed. The Ishii's sawtooth behavior results
from a subtle cancellation of highly non-linear continuum and
ABS's contributions while the normal bound-state contribution
vanishes. For chain hoppings $t_N$ smaller than half of the
superconducting order parameter $\Delta$ we numerically observed
that the dc Josephson current is entirely carried by the ABS's.
This circumstance has been analytically investigated in the limit
of long chains. The condition for the occurrence of the Ishii's
sawtooth behavior is expressed in terms of the microscopic
parameters of the model. We here also point out a limitation of
the WBL approximation, i.e. the independence of the current-phase
relation from $\Delta$.

The ac Josephson regime was studied by applying a constant bias
voltage across the junction and solving numerically the
time-dependent Bogoliubov-deGennes equations for finite leads. We
used the partition-free initial conditions for which the system is
contacted and in equilibrium before an external driving force is
switched on. If the leads are sufficiently long the results of the
time propagation are the same as those of a truly infinite systems
up to a critical time at which finite size effects
appear.\cite{tdspin} Such critical time is, however, large enough
to allow for studying transient responses as well as the ac
Josephson regime setting in after all transient effects have been
washed out. The transient time-scale is dictated by the dwelling
time during which an electron undergoes several Andreev
reflections before being transmitted. By extracting the dc
component of the ac Josephson current we have been able to
reproduce a well-defined subgap structure in the current-voltage
characteristics of a S-S junction. As expected the characteristics
displays kinks at biases $\sim 2\Delta/n$. The time-dependent
approach also permits to perform a spectral decomposition of the
ac current. By Fourier transforming the curve $I_L(t)$ in a proper
time window we computed both the dissipative and non-dissipative
components. Such procedure can be easily generalized to arbitrary
time-dependent fields like, e.g., ac or pulsed biases, at the same
computational cost and provides an alternative approach to
Floquet-based schemes\cite{cuevas1,guo}. We also wish to emphasize
that within the present approach a full microscopic description of
the superconductors is provided, and hence we are able describe
the electron dynamics not only inside the normal region, but also
inside the leads\cite{vanleuw}. This allows us to gain further
information with respect to the WBL approximation. In conclusion
we would like to point out the proposed time-dependent approach is
not limited to 1D electrodes and can be readily generalized to
investigate more realistic superconductor-normal metal interfaces.
In particular it would be interesting to study the case in which
the normal region is two-dimensional, since it has been
experimentally observed\cite{2jos1,2jos2} and theoretically
predicted\cite{kummel2} that in such systems the ac Josephson
current displays a dominant Fourier component at twice the
fundamental Josephson frequency.

\acknowledgments

E.P. is financially supported by Consorzio Nazionale
Interuniversitario per le Scienze Fisiche della Materia.

\appendix

\section{Derivation of the embedding self-energy}
\label{self}

For 1D leads the coupling $V_{q}$ is given in
Eq.(\ref{1dcoupling}) and therefore the retarded embedding
self-energy for lead $\alpha$ in Eq.(\ref{selfen}) reads
\begin{equation}
\underline{\Sigma}^{r}_{\alpha}(\omega)=\frac{2}{\Lambda}
t_{T}^{2}\sum_{q}\sin^{2} (q) \, \sigma_{z} \,
\underline{g}^{r}_{\alpha q} (\omega) \, \sigma_{z} \, .
\label{slf2}
\end{equation}
The retarded Green's function of the uncontacted $\alpha$ lead at
half-filling is given in terms of the $2\times 2$ $q$-dependent
Bogoliubov-deGennes Hamiltonian
\begin{equation}
\underline{H}_{\alpha q} = \left(\begin{array}{cc}
\varepsilon_{q}& \Delta_{\alpha}e^{i\chi_{\alpha}}
\\\Delta_{\alpha}e^{-i\chi_{\alpha}}&-\varepsilon_{q}
\end{array}\right)
\end{equation}
as
\begin{equation}
\underline{g}^{r}_{\alpha q} = \frac{1}{\omega
-\underline{H}_{\alpha q}+ i\eta } .
\end{equation}
Introducing the eigenvectors
\begin{equation}
|\psi^{\pm}_{\alpha q} \rangle = \left(\begin{array}{c}
\sqrt{\frac{1}{2}\left( 1 + \frac{\varepsilon_{q}
}{\xi^{\pm}_{\alpha
q}} \right)} \\
\pm e^{-i\chi_{\alpha}} \sqrt{\frac{1}{2}\left( 1 -
\frac{\varepsilon_{q} }{\xi^{\pm}_{\alpha q}} \right)}
\end{array}\right)
\end{equation}
of $\underline{H}_{\alpha q}$ with eigenvalues $\xi^{\pm}_{\alpha
q}= \pm \sqrt{\varepsilon_{q}^{2}+\Delta_{\alpha}^{2}}$ and taking
the limit $\Lambda \to \infty$ Eq.(\ref{slf2}) becomes
\begin{equation}
\underline{\Sigma}^{r}_{\alpha}(\omega)=2t_{T}^{2}\int_{0}^{\pi}
\frac{dq}{\pi} \sin(q)^{2} \, \sum_{\nu = \pm} \sigma_{z} \,
\frac{|\psi_{\alpha q}^{\nu} \rangle \langle \psi_{\alpha q}^{\nu}
| }{\omega - \xi^{\pm}_{\alpha q} +i \eta } \,\sigma_{z} \,.
\label{slf3}
\end{equation}
The integral can be computed analytically to yield
\begin{equation}
\underline{\Sigma}^{r}_{\alpha}(\omega)=\left(
\begin{array}{cc}m_{\alpha}(\omega+i\eta)&
\tilde{\Delta}_{\alpha}(\omega+i\eta) e^{i\chi_{\alpha}}
\\\tilde{\Delta}_{\alpha}(\omega+i\eta)
e^{-i\chi_{\alpha}}&m_{\alpha}(\omega+i\eta)
\end{array}\right) \, ,
\end{equation}
where
\begin{eqnarray}
m_{\alpha}(z) &=& z \frac{t_{T}^{2}}{2t_{S}^{2}}
\frac{\sqrt{\Delta_{\alpha}^{2}-z^{2}}-\sqrt{\Delta_{\alpha}^{2}-
z^{2}+4t_{S}^{2}}}{\sqrt{\Delta_{\alpha}^{2}-z^{2}}} \, ,
\nonumber \\
\tilde{\Delta}_{\alpha}(z)&=&\Delta_{\alpha}
\frac{t_{T}^{2}}{2t_{S}^{2}} \frac{\sqrt{
z^{2}-\Delta_{\alpha}^{2}-4t_{S}^{2}}-\sqrt{z^{2}-
\Delta_{\alpha}^{2}}}{\sqrt{z^{2}-\Delta_{\alpha}^{2}}} \,,
\label{embedding}
\end{eqnarray}
with $z$ is a complex frequency.

The other relevant components of the Nambu self-energy are easily
obtained starting from the retarded one:
\begin{eqnarray}
\underline{\Sigma}^{a}_{\alpha}(\omega)&=&\left[
\underline{\Sigma}^{r}_{\alpha}(\omega)\right]^{\dagger} \,,
\nonumber \\ \underline{\Sigma}^{<}_{\alpha}(\omega)&=&-f(\omega)
[\underline{\Sigma}^{r}_{\alpha}(\omega)-
\underline{\Sigma}^{a}_{\alpha}(\omega)] \, ,
\end{eqnarray}
where $f$ is the Fermi distribution function.

\section{Calculation of the bound states}
\label{boundstates}

In this Appendix we solve the  eigenvalue problem in
Eq.(\ref{secular}). We use the following ansatz\cite{affleck2} for
the eigenstate amplitudes $\psi_{k}(j)$ on the $j$-th site of the
normal region
\begin{equation}
\psi_{k}(j)=\left(\begin{array}{c} u_{k}(j)
\\v_{k } (j)
\end{array}\right) = \left(\begin{array}{c} A_{k}e^{ikj}+B_{k}e^{-ikj}
\\\ (-1)^{j}(C_{k}e^{ikj}+D_{k}e^{-ikj})
\end{array}\right) \, ,
\end{equation}
where $j=1,...M$. Due to the symmetry of the problem the
wavefunction must be chosen so as to fulfill the condition
\begin{equation}
|u_{k}(1)|=|u_{k}(M)| \,,
\label{kall}
\end{equation}
which is equivalent to $|v_{k}(1)|=|v_{k}(M)|$. The above
condition provides an equation for the allowed wavevectors $k$,
similarly to the case of normal open chains. In the following we
specialize to even $M$. The case of odd $M$ is similar and does
not introduce  extra complications. We first observe that due to
the choice $\Delta_{L}=\Delta_{R}$ the effective Hamiltonian in
Eq.(\ref{heff}) is invariant under the transformation $T$:
$c_{j\uparrow} \rightarrow (-1)^{j}\tilde{c}_{M+1-j \downarrow}$
and $\tilde{c}_{j\downarrow} \rightarrow (-1)^{j} c_{M+1-j
\uparrow}$. It is straightforward to realize that $T^{2}=-1$ and
hence the wavefunctions obey the symmetry constraint $v_{k}(M)=i
\nu u_{k}(1)$ where $\nu = \pm$ is a parity index. By applying the
Schrodinger equation to sites 1,2 for spin $\uparrow$ and to sites
$M,M-1$ for spin $\downarrow$ and exploiting the above symmetry
constraint we obtain the following linear systems for the
coefficients $A_{k},B_{k},C_{k},D_{k}$
\begin{widetext}
\begin{equation}
\left(\begin{array}{llll}
 e^{ik}&  e^{-ik} &0 &0 \\
t_{N} e^{2ik}& t_{N} e^{-2ik}& -\tilde{\Delta}_{k}e^{-i\chi
/2}e^{ik}
& -\tilde{\Delta}_{k}e^{-i\chi /2}e^{-ik}\\
0& 0&  e^{ikM}&  e^{-ikM}  \\
-\tilde{\Delta}_{k}e^{-i\chi /2}e^{ikM}&
-\tilde{\Delta}_{k}e^{-i\chi /2}e^{-ikM}& -t_{N} e^{ik(M-1)}&
-t_{N} e^{-ik(M-1)} \end{array}\right)
\left(\begin{array}{c} A_{k} \\
B_{k} \\ C_{k} \\ D_{k}
\end{array}\right)
 =
\left(\begin{array}{l} 1 \\ 2t_{N}\cos(k) -m_{k} \\
i \nu  \\
-i \nu(2t_{N}\cos(k) -m_{k})
\end{array}\right) \, ,
\end{equation}
\end{widetext}
where we have chosen $u_{k}(1)=A_{k}e^{ik}+B_{k}e^{-ik}=1$. Indeed
the proper normalization factor of the Andreev bound state
wavefunction is inessential to the calculation of the bound state
energy. The solution of the above system provides the
$k$-dependent coefficients
$$
A_{k}= \frac{1}{\Omega_{k}}\left[e^{i(\frac{\chi}{2}+k M)}(m_{k}-
t_{N}e^{i k}) -i\nu \tilde {\Delta}_{k} e^{ik}\right],
$$
$$
B_{k}=
\frac{e^{ik(M+1)}}{\Omega_{k}}\left[e^{i\frac{\chi}{2}}(t_{N}-m_{k}e^{ik})
+i \nu \tilde {\Delta}_{k} e^{ikM}\right],
$$
$$
C_{k}= \frac{1}{\Omega_{k}}\left[i \nu e^{i\frac{\chi}{2}}(t_{N} -
m_{k}e^{ik}) -\tilde {\Delta}_{k} e^{ikM}\right],
$$
$$
D_{k}= \frac{e^{ik(M+1)}}{\Omega_{k}}\left[i \nu
e^{i(\frac{\chi}{2}+k M)} (m_{k} - t_{N} e^{ik})+\tilde
{\Delta}_{k} e^{ik}\right],
$$
where
\begin{equation}
\Omega_{k}=t_{N}e^{i(\frac{\chi}{2}+kM)}(e^{2ik}-1)-i \nu \tilde
{\Delta}_{k} (e^{2ik}-e^{2ikM}) \, .
\end{equation}
Inserting the above solution in Eq.(\ref{kall}) and taking into
account the normalization condition $u_{k}=1$ one finds the
following equation for the allowed values of $k$
\begin{eqnarray}
0&=&2\left[t_{N}\sin(kM)-m_{k}\sin
k(M-1)\right]^{2}-t_{N}^{2}-\tilde{\Delta}_{k}^{2} \nonumber \\
&+&t_{N}^{2}\cos (2k) + \tilde{\Delta}_{k}^{2} \cos
2k(M-1)\nonumber \\
&-& \nu 4t_{N} \tilde{\Delta}_{k} \cos (\chi/2) \sin (k) \sin
k(M-1) \, .
\end{eqnarray}
Isolating the last term and squaring, the dependence on $\nu$
disappears and we end up with an equation valid for both parities
\begin{eqnarray}
0&=& t_{N}^{4} \sin ^{2}k(M+1)
+(-1)^{M}2t_{N}^{2}\tilde{\Delta}_{k}^{2} \cos \chi \sin ^{2}k
\nonumber \\
&-&2t_{N}\tilde{\Delta}_{k}^{2} \sin^{2}(kM) +
\tilde{\Delta}_{k}^{4} \sin^{2}k(M-1) \nonumber
\\
&-&m_{k}^{2}[2t_{N}\sin (kM)-m_{k}\sin k (M-1)]^{2} \,,
\label{andreev}
\end{eqnarray}
which coincides with Eq.(\ref{andreevt}). The bound states
eigenenergies $E^{(k)}= 2t_{N}\cos k$ are obtained by solving
Eq.(\ref{andreev}) numerically and retaining only the values of
$k$ such that $|E^{(k)} | < \Delta$ (Andreev bound states) and
$|E^{(k)}|>\sqrt{4t_{S}^{2}+\Delta^{2}}$ (normal bound states). We
notice that the wavevectors $k$ for the ABS's are real valued
while for normal bound states are in general complex.  Indeed for
$|t_{N}| <\frac{1}{2} \sqrt{4t_{S}^{2}+\Delta^{2}}$ the energy
$E^{(k)}$ lies below/above the continuum only for complex $k$.


\end{document}